\documentclass[preprint,authoryear,12pt]{elsarticle}
\def\mbh{$M_{\rm BH}$\/}

\def\mbh{$M_{\rm BH}$\/}
\def\lledd{$L/L_{\rm Edd}$}
\def\ne{$n_{\rm e}$\/}
\def\nh{$n_{\rm H}$\/}
\def\nc{$N_{\rm c}$\/}

\def\msol{M$_\odot$\/}

\def\rg{$R_{\rm g}$\/}
\def\ltsima{$\; \buildrel < \over \sim \;$}
\def\simlt{\lower.5ex\hbox{\ltsima}}            % < over MMM
\def\gtsima{$\; \buildrel > \over \sim \;$}

\def\simgt{\lower.5ex\hbox{\gtsima}}            % > over MMM

\def\civ{{\sc{Civ}}$\lambda$1549\/}
\def\civnc{{\sc{Civ}}$\lambda$1549$_{\rm NC}$\/}
\def\civbc{{\sc{Civ}}$\lambda$1549$_{\rm BC}$\/}

\def\cm3{cm$^{-3}$\/}
\def\hb{{\sc{H}}$\beta$\/}

\def\hbbc{{\sc{H}}$\beta_{\rm BC}$\/}

\def\mgii{{Mg\sc{ii}}$\lambda$2800\/}

\def\oiiiopt{{\sc{[Oiii]}}$\lambda\lambda$4959,5007\/}
\def\o4363{{\sc{[Oiii]}}$\lambda$4363\/}

\def\siiii{{Si}{\sc iii}]$\lambda$1892\/}

\def\feii{{Fe\sc{ii}}\/}
\def\fe{{\sc{Fe}}\/}

\def\fe76087{{\sc [Fe vii]}$\lambda$6087\/}
\def\oiii{{\sc [Oiii]}$\lambda$5007}
\def\siiv{{Si{\sc iv}$\lambda$1397}}
\def\aliii{{Al{\sc iii}$\lambda$1860}}
\def\oiv{{O{\sc iv}]$\lambda$1402}}

\def\kms{km~s$^{-1}$}

\def\ergss{erg\, s$^{-1}$\/}

\def\hi{H{\sc i}\/}

\def\rb{$r_{\rm BLR}$\/}

\def\mbulge{$M_{\mathrm{bulge}}$}

\def\sii{[S{\sc ii}]$\lambda\lambda$6716,5631}
\def\oii{[O{\sc ii}]$\lambda$3727}

\usepackage{graphicx}
\usepackage{epsfig}

\usepackage{amssymb}

\journal{New Astronomy Reviews}

\begin{document}

\begin{frontmatter}



\title{Estimating Black Hole Masses in Quasars \\ Using Broad Optical and UV Emission Lines}


\author{Paola Marziani and Jack Sulentic}

\address{INAF, Osservatorio Astronomico di Padova, Italia \& Instituto de Astrofis\'{\i}ca de Andaluc\'{\i}a, CSIC, Espa{\~n}a}

\begin{abstract}

We review past work using broad emission lines as virial estimators of black hole masses in quasars. Basically one requires estimates of the emitting region radius and virial velocity dispersion to obtain black hole masses. The three major ways to estimate the broad-line emitting region (BLR) radius involve: (1) direct reverberation mapping, (2) derivation of BLR radius for larger samples  using the radius-luminosity correlation derived from reverberation measures, and (3) estimates of BLR radius using the definition of the ionization parameter solved for BLR radius (photoionization method). At low redshift ($z \simlt$ 0.7) FWHM H$\beta$ serves as the most widely used estimator of virial velocity dispersion. FWHM H$\beta$ can provide estimates for tens of thousands of quasars out to $z \approx$ 3.8 (IR spectroscopy beyond $z \approx$ 1).  A new photoionization method also shows promise for providing many reasonable estimates of BLR radius via high S/N IR spectroscopy of the UV region 1300 -- 2000 \AA.  FWHM \mgii\ can serve as a surrogate for FWHM H$\beta$ in the range $0.4 \simlt z \simlt 6.5$  while \civ\ is affected by broadening due to non-virial motions and best avoided (i.e. there is no clear conversion factor between FWHM \hb\ and FWHM \civ). Most quasars yield mass estimates in the range   7 $\simlt \log M_\mathrm{BH} \simlt $ 9.7. There is no strong evidence for values above 10.0  and there may be evidence for a turnover in the maximum black hole mass near $z  \approx 5$. 
\end{abstract}


\end{frontmatter}
%\tableofcontents
\section{Estimation of Black Hole Masses in Quasars: an Introduction \label{mass}}
\label{intro}

The existence of supermassive compact objects in galactic nuclei was proposed to solve  the energy problem raised by the discovery of radio galaxies \cite{hoylefowler63}.  They estimated that the mass of such an object had to be of the order of 10$^8$\ solar masses. \citet{hoylefowler63} showed that nuclear reactions were insufficient to provide the energy and that the energy source had to be gravitational collapse
\citep[see also ][]{salpeter64,zeldovichnovikov65}. The discovery of quasars \citep{Greensteinschmidt64} and of rapid optical variability \citep{smithhoffeit65} exacerbated the problem because, at their redshift distances,  many quasars apparently involve sources emitting 10$^3 $\ times  the luminosity of an $L^{*}$\  galaxy within a volume much less than a parsec of diameter. Estimation of the masses for these black holes is therefore at the center of quasar astrophysics and cosmology.

The black hole mass of quasars is a fundamental parameter that relates to the evolutionary stage of quasars and of the accretion processes occurring within them. An estimate of black hole mass (\mbh) allows one to assess the role of gravitational forces in the dynamics of the region surrounding the black hole.  The  power output of quasars is directly proportional to $M_{\mathrm{BH}}$. There is much debate over  how the evolution  of quasar energetic output might affect the development and structure of the host galaxy, as well as larger scale  structure formation and, at very high redshifts,  the re-ionization of the Universe \citep[e.g., ][]{fan10}. 

Early estimates of \mbh\ were based on the apparent similarity of quasar spectra and their line widths. The size of the emitting region can be written as:

\begin{equation}
\label{eq:dibai}
\Delta r = \left ( \frac{3 L_\mathrm{line}}{4 \pi f_\mathrm{f} \epsilon_\mathrm{line}} \right )^{\frac{1}{3}}
\end{equation}

where $L_\mathrm{line}$\ is the line (typically \hb) luminosity, $f_\mathrm{f}$\ is the filling factor and $\epsilon_\mathrm{line}$\ is the line emissivity. This expression was used to derive  \mbh\ by considering the nucleus as a bound system and line broadening due to gas cloud motions  \citep{dibai77,dibai84,wandelyahil85}. It is interesting to point out that an analogous argument (coupled with the relatively short timescale of observed continuum variations) was used  immediately  after the discovery of quasars to suggest a small emitting region size and large mass.

The similarity of quasar spectra implies a roughly constant ionization parameter $U$\ or a constant product $U$\ times the electron density \ne\    \citep{baldwinnetzer78,davidsonnetzer79}, where

\begin{equation}
\label{eq:u}
U = \frac {\int_{\nu_0}^{+\infty}  \frac{L_\nu} {h\nu} d\nu} {4\pi n_\mathrm{e} c r^2}.
\end{equation}

Here $L_{\nu}$\ is the specific luminosity per unit frequency, $h$\ is  the Planck constant, $\nu_{0}$\ the Rydberg frequency, $c$ the speed of light, and $r$\ can be interpreted  as the distance between the central source of ionizing radiation and the line emitting region.   So basically \rb\ is $\propto L^{1/2}$ using rather order-of-magnitude considerations and the assumption of constant $U$\  pointing towards  large masses increasing
with source luminosity.

 \begin{figure}
 \epsfxsize=15cm
 \epsfbox{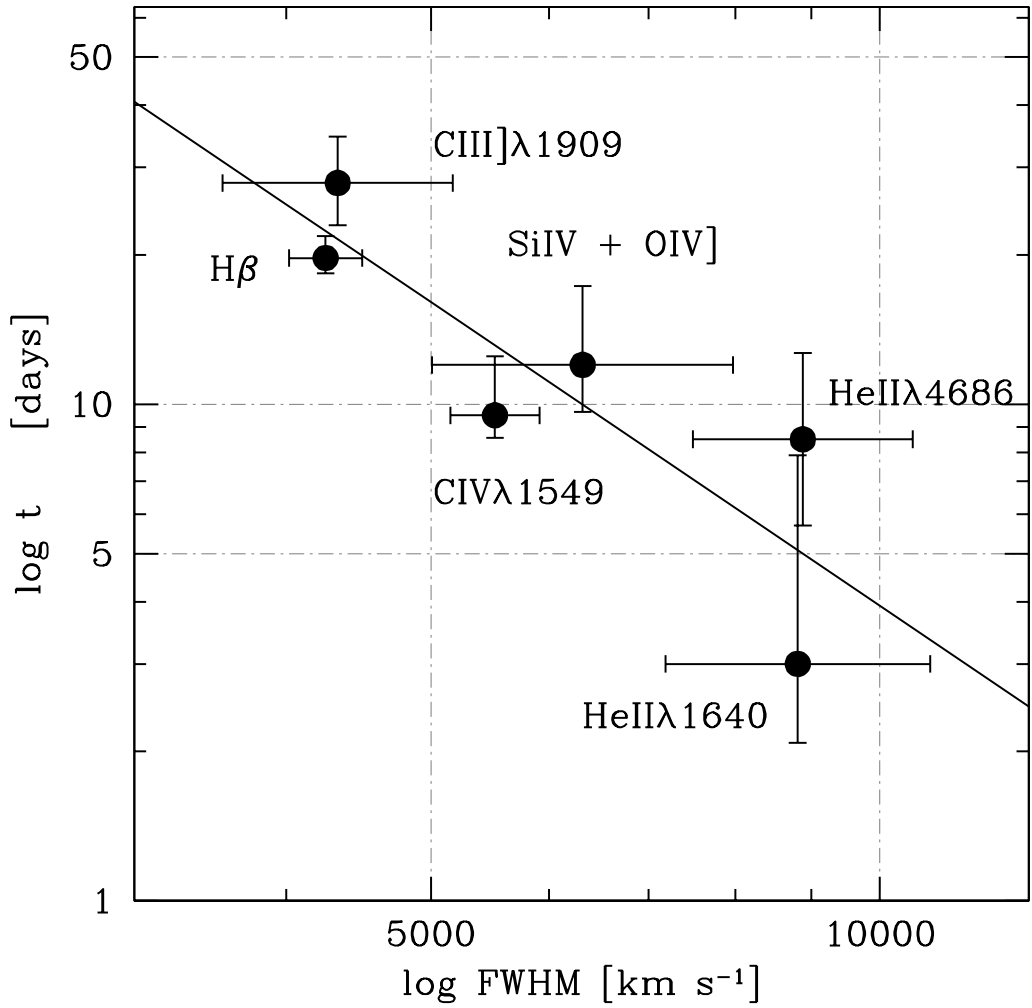}
 \caption[]{Test of virial broadening on several emission lines of NGC 5548. Adapted from \cite{petersonwandel99}. The line shows an unweighted least square fit. } \label{fig:virial}
 \end{figure}

% \vspace{8cm}

\section{The Virial Assumption}

The virial mass is defined as:

\begin{equation}
M_{\rm BH } = f \frac{r \delta v_\mathrm{r}^2 }{ G},
\label{eq:mass}
\end{equation}

where $f$ is a factor  dependent on geometry of the emitting region, $r$ is the distance of line emitting gas  from the central black hole, $\delta v_\mathrm{r}$\ is the line broadening due to virial motions, and $G$\ is the gravitational constant. All methods based on broad optical and UV lines (e.g. photoionization, reverberation) come down to estimating  the broad line region (BLR) distance from the central continuum source (hereafter indicated as the BLR radius \rb) and combining it with a broad line width to derive \mbh.

The term ``Keplerian'' is used to indicate the particular case of a gravitationally bound system where the virial theorem applies. Kepler's third law is appropriate when mass is mainly concentrated in a single body and  the self-interaction gravitational potential of any remaining mass can be neglected; in the case of quasars, \mbh$\gg$ mass of the line emitting gas (i.e., not unlike the solar system if solar and planetary masses are compared).  Therefore, the terms virial and Keplerian will be used synonymously in the quasar context. 

The fundamental question is whether the virial assumption is correct for the line being used. Evidence in support of virialization come from  early velocity resolved reverberation mapping studies \citep{gaskell88,koratkargaskell89,koratkargaskell91} that excluded outflow as the broadening source in at least low-ionization lines. Circumstantial evidence for virial motion involves emission line profiles of NLSy1-like sources which are relatively symmetric and smooth \citep{marzianietal03a}.

  If the virial assumption is valid then the velocity field is Keplerian and  line broadening should anti-correlate with the time lag of different lines \citep{petersonwandel99,petersonetal04,krolik01}.  Sufficient data to test the width-distance relation are  available  for NGC 5548. Figure \ref{fig:virial} shows a trend ($r \propto \delta v_\mathrm{r}^{2}$) consistent with dependence of time delay on line width. The delays of 6 emission lines (including the blend \oiv+\siiv) are plotted versus FWHM measures. Data come from the International AGN watch carried out in 1989. An unweighted least squares fit yields a slope of $\alpha = 2.01 \pm 0.60$ which is consistent with the value  expected for a Keplerian velocity field. A weighted least squares fit that takes into account errors on both axes yields a slope $2.60 \pm 0.60$ indicating a strong increase in FWHM with decreasing distance. 

Equation \ref{eq:mass} is deceptively simple.  Every term is problematic: \begin{enumerate} 
\item The geometric factor $f$ allows us to convert an observed FWHM value into a virial velocity. $f$ is poorly known because BLR geometry and effects of line-of-sight orientation are poorly known.  
\item The distance $r$  is highly uncertain because the BLR is not spatially resolved in any source.   
\item The appropriateness  of  $\delta v$\ (virial broadening) rests on the assumption that lines are Doppler  broadened reflecting motions of the emitting gas, and not by scattering (as considered by \citealt{mathis70,kallmankrolik86,gaskellgoosmann08}).  To estimate $\delta v$  one must determine which  line or line component (if any)  is broadened by the motions of virialized gas. Line profile  shifts relative to the quasar rest frame warn against indiscriminate use of line profile widths to estimate virial broadening. 
\end{enumerate}

\section{The factor $f$}

Our uncertainty about the geometric parameter likely represents a roadblock to estimation of  black hole masses more accurate than a factor of  3 -- 5. If FWHM of a broad line is chosen as an estimator of $\delta v_\mathrm{r}$  and the velocity distribution is isotropic then the square of the velocity module is $v^{2} = 3 v_\mathrm{r}^{2}$ implying a geometric factor $f \approx \sqrt{3}/2$.   In Eq. \ref{eq:mass} our lack of knowledge about the geometry and dynamics of the BLR is concentrated in $f$. In other words, an understanding of the dynamics and geometry of the BLR is directly related to the determination of \mbh. Considerable recent discussion and modeling have been directed towards estimating $f$\ and how it might depend on changing physical conditions \citep{collinetal06,onkenetal04,netzermarziani10}.  Radiation pressure forces and orientation effects make it possible that it is significantly ($\times$3 in \citealt{collinetal06}) and systematically different in Pop.  A (FWHM \hb $\simlt$4000 \kms) and Pop. B sources (FWHM \hb $\simgt$4000 \kms). Pop. A and B show striking differences in the \hb\ and \civ\ profiles and in several other properties \citep{sulenticetal07}. Although their interpretation in terms of BLR structure is not yet clear, further discussion in \S \ref{popab} will enlighten  the importance of separating quasars into these two populations.

\subsection{Dynamical effects}

Given the strong radiation field to which gas is exposed in quasars it seems unlikely that gravity is the only important force. The relative balance of gravitational and radiation forces in a source of fixed \mbh\ points toward a role for Eddington ratio. The ratio between radiative and gravitational acceleration in an optically thick, Compton-thin, medium can be written as

\begin{equation}
r_\mathrm{a} =  \frac{a_{\mathrm{rad} } }{a_{\mathrm{grav} } } \approx 7.2 \frac{L_\mathrm{bol}}{L_\mathrm{Edd}}  N_\mathrm{c,23}^{-1} \label{eq:radgrav}
\end{equation}
where $L_\mathrm{bol}$\ and $L_\mathrm{Edd}$ are  the bolometric and  Eddington luminosities,  $N_\mathrm{c,23}$ is the hydrogen column density in units of 10$^{23}$ cm$^{-2}$. If $r_\mathrm{a} >1$ \ radiative acceleration dominates \citep[cf.][]{netzermarziani10,marzianietal10}. 

Very high column density clouds will be unaffected by radiative forces while lower column densities will be  affected even at moderate Eddington ratios. A more refined dynamical treatment should consider the effects of radiative forces on  a system of clouds moving in pressure balance with an external medium. 

%Since the gas density in individual clouds is dependent  on the radial coordinate, also the column density -- and hence the relative relevance of radiation pressure and gravitation force -- will depend on cloud location. 
 
\begin{center}
\begin{figure}[ht!]
 \epsfxsize=13cm
 \epsfbox{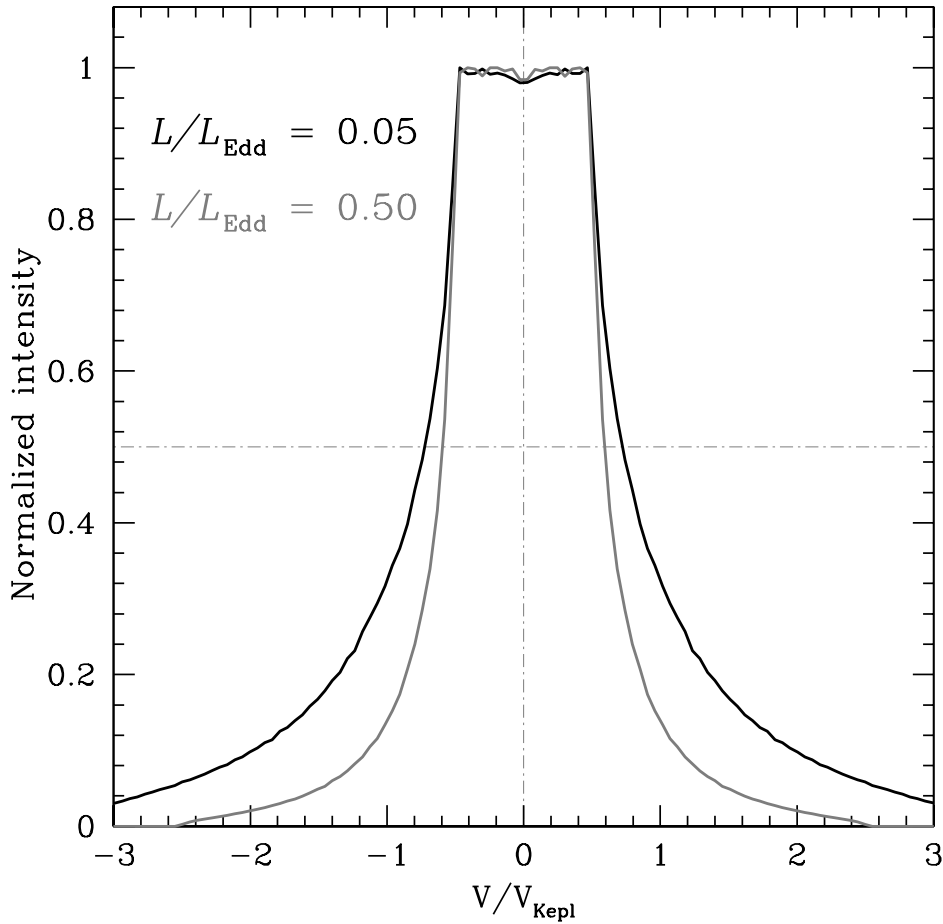}
 \caption[]{Effects of radiation pressure on line profiles computed for a spherical distribution of clouds in pressure equilibrium following \citet{netzermarziani10}. Model parameters are kept the same save for the Eddington ratio that has been increased by a factor 10, from \lledd = 0.05 to \lledd = 0.5 . The dot-dashed lines mark the rest-frame origin and the half maximum intensity level. }
\label{fig:radpr} 
\end{figure}
\end{center}

This approach indicates that the effect of radiation pressure significantly alters the equilibrium of a system of clouds: the emissivity weighted radius will increase as orbits  become more elliptical and hence the cloud will spend a longer time at larger distances from the central continuum source. The value of $f$\ will therefore depend on Eddington ratio. At the same time the line will become narrower as radiation pressure increases. Fig. \ref{fig:radpr}  shows two \hb\ profiles computed for confined clouds moving   in pressure equilibrium with an external medium assuming a spherical distribution of orbits  \citep{netzermarziani10}. Eddington ratio that has been set to \lledd = 0.05 (at this low \lledd\  the effect of radiative forces is small)  and to  \lledd = 0.5 (appropriate for NLSy1-like Pop. A sources). The effect on FWHM and on the virial product \rb FWHM$^{2}$\ is rather modest, and the derived $f$  increases from $\approx $0.76 to $\approx$ 1.08 along with \lledd. How will these results affect single-epoch \mbh\ estimates in a large sample? The dependence of \rb\ on Eddington ratio will be a source of scatter in the \rb-$L$\ correlation. If this is taken into account we find an effect of less than a factor of  2 for \mbh\ estimates over 4dex in luminosity \citep{netzermarziani10}.  

%If this is taken into account the effect of 4 orders of magnitude in luminosity on \mbh\ is relatively small, less than a factor 2. 

Previous work \citep{marconietal08,marconietal09} accounted for radiation pressure effects by adding an \mbh\ term proportional to the ratio $L$/\nc\ (where \nc\ was supposed to have a log-normal distribution with average $\log$\nc = 23).  In this case corrected \mbh\  can be a factor of even $\simgt$10  higher than  values obtained using the virial formula \citep{gaskell96}. The implied  masses of the most luminous quasars that would have been $\simgt 10^{10}$\msol, possibly reintroducing what we have called the ``maximum mass problem'' \citep{netzer03}: the most massive black holes at any redshift cannot exceed by a large factor the maximum mass of the (mostly quiescent) black holes at low $z$, around $5\cdot 10^{9}$ \msol\ (the mass of the black hole in Messier 87; \citealt{macchettoetal97}). In other words, black holes with masses above  $5\cdot 10^{9}$\ \msol\ should be extremely rare (\citealt{sulenticetal06}  and references therein). The \citet{marconietal09} results are probably incorrect because they neglect the likelihood that the column density of bound clouds is dependent on Eddington ratio and hence on luminosity for a given mass. %On the other hand, the treatment of \citet{netzermarziani10} rests on . 

\subsection{Estimating $f$ empirically: normalization}
\label{norm}

The zone of gravitational influence for the black hole has been resolved in several nearby galaxies. The most reliable methods involve observation of H$_{2}$O masers \citep{miyoshietal95,greeneetal10,kuoetal11} or, at a second stance, space-based long-slit spectra of optical emission lines \citep[e.g.,][]{ferrareseetal96,macchettoetal97,barthetal01,capettietal05}. Both methods spatially resolve the velocity field and yield a radial velocity ``cusp'' indicative of distances where the velocity field is governed by gravity of the black hole. Mass values obtained from gas dynamics are considered less reliable since the line emitting gas might be more affected by non-circular motions \citep[e.g., ][]{barthetal01,shapiroetal06}.  

Black hole masses computed from resolved velocity fields correlate with stellar velocity dispersion measures in the bulges of host galaxies \citep{ferraresemerritt00,gebhardtetal00}. 
The \mbh\  -- $\sigma_{\star}$\ correlation obtained for velocity resolved galaxies has been used to estimate $f$\ which can then be used for computation of quasar BH masses obtained from reverberation mapping derivations of \rb. One basically overlaps the \mbh\  -- $\sigma_{\star}$\ relations for non-active galaxies and AGN where  $\sigma_{\star}$ has been estimated, usually from the IR Calcium triplet \citep{onkenetal04,collinetal06,gultekinetal09,wooetal10,grahametal11}. 

The derived $f_{\sigma} \approx 5.5$ of \citet[][where $f_{\sigma}$\ refers to the $f$ value when the velocity dispersion is used as virial broadening estimator]{onkenetal04} was reputed a valid estimate but a recent study now suggests $f_{\sigma} \approx 2.9$ \citep{grahametal11}, halving the \mbh\  values. Considering that the \mbh\  -- $\sigma_{\star}$\ relation for nearby galaxies shows significant scatter, results are subject to substantial uncertainty.  In addition,  there are two fundamental caveats: 
 
 \begin{enumerate}
\item  \mbh\  -- $\sigma_{\star}$ correlation of nearby galaxies could be a selection effect \citep{kormendy93}: only the most massive black holes can be resolved for a given bulge mass. 
%It is still possible that a range of black hole masses is present for a given bulge mass. While probably valid for large masses, 
The relation between \mbh\ and  $\sigma_\star$\ should be taken with special care in the lower \mbh\ range. This issue is farther discussed in \S \ref{bulge}.

\item It is unlikely that a single $f$\ value is valid. $f$\ appears to depend on the width of the \hb\ line or, almost equivalently, on its shape.  \citet{collinetal06} separated the reverberation sample in two populations: the Pop. A and B of \citet{sulenticetal00a} and Pop. 1 and 2 according to a profile shape criterion.  This difference reflects   different geometries  that should also lead  to  different responses to continuum changes. This most interesting result of \citet{collinetal06} should be taken into account in any eventual attempt to derive $f$.  
\end{enumerate}

%based on the ratio between FWHM and $\sigma$: FWHM/$\sigma < $ 2.35 sources were classified Pop. 1, and Pop. 2 otherwise. FWHM/$\sigma$\ increased from spikier to boxier objects, with FWHM/$\sigma \approx$  2.35 being the value of a Gaussian. There is a rough correspondence between Pop. A and Pop. 1, and Pop. B and Pop. 2 since Pop. A spectra show \hb\ with a Lorentzian-like profile \citep{sulenticetal02} and Pop. B show double-Gaussian, broader \hb\ profile. Derived $f$ values are $f \approx 2$\ for Pop. A sources, and $f \approx 0.5 - 1.0$. 

\subsection{Orientation Effects}

\begin{center}
\begin{figure}[ht!]
 \epsfxsize=13cm
 \epsfbox{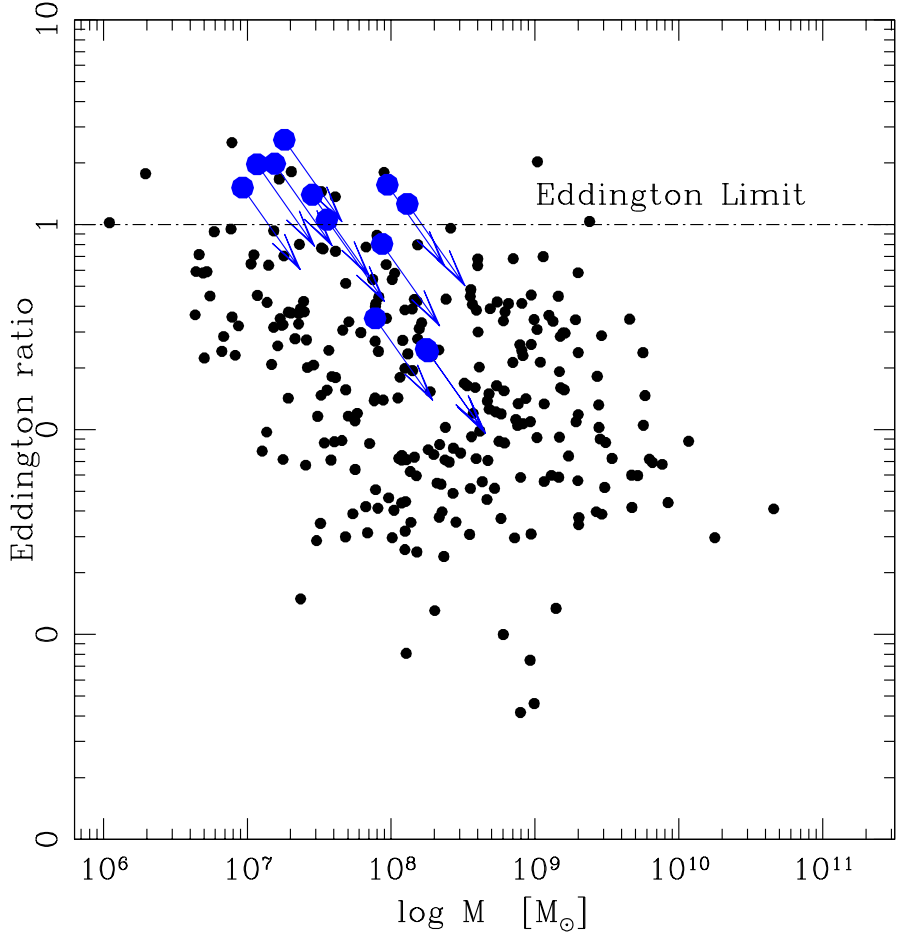}
 \caption[]{Effects of orientation on the low-$z$ sample of \citet{marzianietal03b}. Blue outliers (large blue spots) are indicated along with  other sources in our sample (filled circles) in the plane \lledd\ vs. \mbh. The largest circle indicates the position of PKS 0736+01, a radio-loud blue outlier.  Arrows indicate displacements of blue outliers if we apply an orientation correction to their masses.  }\label{fig:orient} 
 \end{figure}
\end{center}

\label{orient}
Orbits  of BLR clouds are unlikely to be oriented randomly resulting in a strong dependence of \mbh\ on orientation. The $f$\ value derived by \citet{collinetal06} indicates that low-ionization line (LIL) emission in Pop. A sources may arise in a flattened configuration.  Other lines of evidence suggesting a flattened gas distribution have been reported since the 1970s (they have been recently reviewed by \citealt{gaskell09b}). Sources viewed at high inclination will then be measured with too small FWHM leading to  systematic underestimation of \mbh. 

Measures of both radio-quiet and radio-loud quasars likely suffer from inclination effects with the latter offering a rather straightforward way of estimating its amplitude. One can compare average FWHM \hb\ values for core-dominated  and lobe-dominated sources. If the radio jet  is oriented perpendicular to a flattened distribution of line emitting clouds then  core- and lobe-dominated represent sources where the disk is viewed near and far from face-on respectively. Lobe-dominated sources are found to be broader by a factor 2 \citep{mileymiller79,sulenticetal03,zamfiretal08}. The assumption of an isotropic velocity component combined with a rotational one  roughly envelops the distribution of radio-loud sources in the plane $R$ vs FWHM \hb\ parameter plane \citep{willsbrowne86}.  In a small subsample of radio-loud sources with detected superluminal motion the apparent speed can be used to retrieve the angle $\theta$\ between the line-of-sight and the relativistically moving plasma, once an estimate of the Lorentz factor $\gamma$ is obtained in the framework of the synchrotron self-Compton model of X-ray and radio emission \citep{rokakietal03,sulenticetal03}. A diagram of $\theta$ vs. FWHM confirms a factor $\simgt$ 2 broadening for the sources observed at larger inclination. An effect of similar amplitude is expected for radio-quiet sources, although attempts have yet to yield a convincing orientation indicator (see however the recent work of \citealt{boroson11}). Considering that $\theta$
may be in the range $few \, degrees\, \simlt \theta \simlt 45^\circ$,  and distributed randomly, we can have an average $<\theta> \approx 30^\circ$; this means that, ignoring orientation effects leads to
 an underestimate of the mass for almost all sources, up to a factor $\sim $ 10 if the source is observed almost pole-on and if the BLR is strongly flattened. The effect is less (a factor $\approx 5$) in the case of Pop. B sources where a vertical component of motion (possibly due to turbulence)  is more strongly affecting the line width \citep[e.g., ][]{gaskell09b}. Even if pole-on sources are rarest in a randomly-oriented sample,
 errors of this amplitude may dramatically increase the scatter in large samples and influence inferences on
 Eddington ratio \citep{jarvismclure06}. Fig. \ref{fig:orient} shows the rare population of ``blue outliers'' i.e., Pop. A sources showing a large \oiiiopt\ blueshift  \citep{zamanovetal02} and believed to be oriented almost pole-on \citep{boroson11}. If broad \hb\ is emitted in a highly flattened configuration then they will appear systematically undermassive and hence super-Eddington.  An orientation correction of 0.4 (in logarithm) to  estimated masses of the blue outliers   moves them in the sub-Eddington regime of \lledd\ vs \mbh\ diagram.

 \begin{figure}[ht!]
\epsfxsize=15cm
 \epsfbox{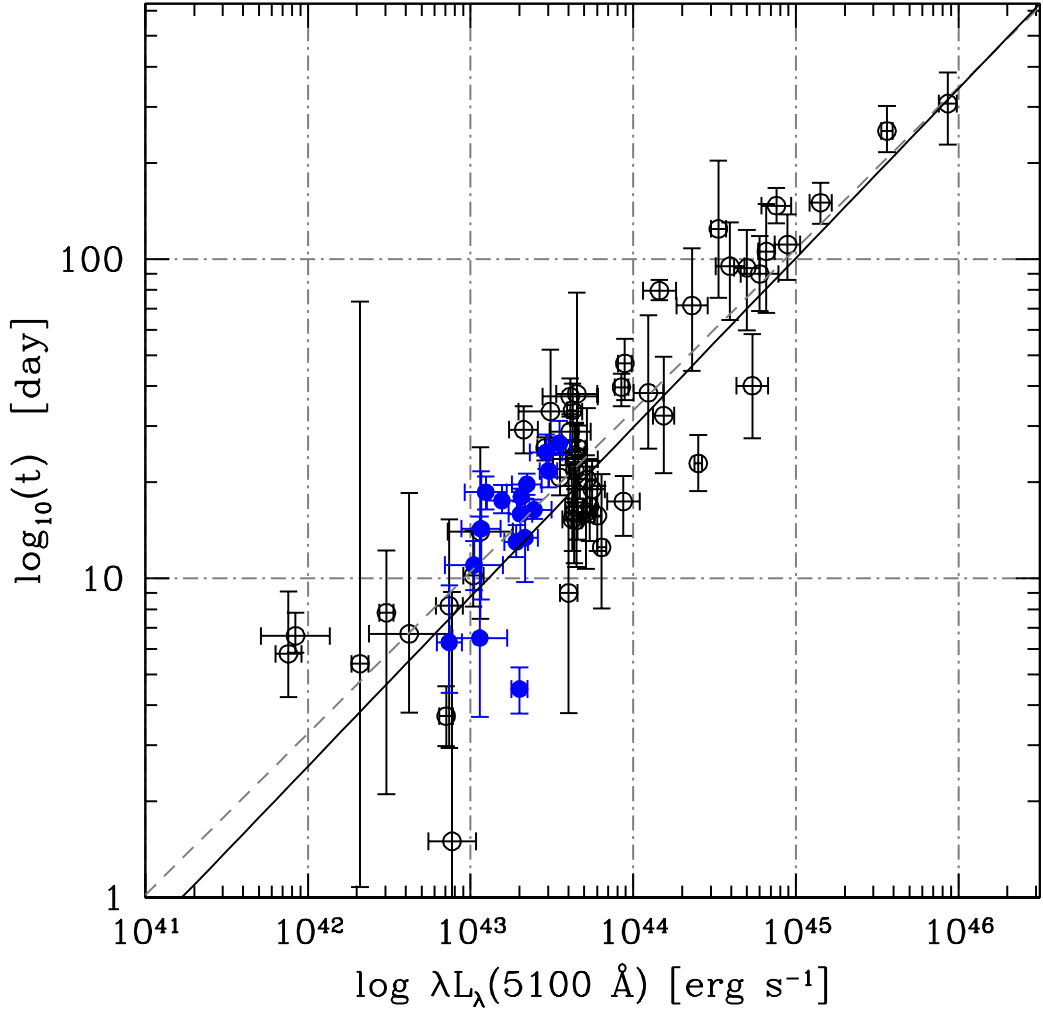}
\caption[]{Correlation between time delay in units of day and luminosity at 5100 \AA\ in units of \ergss. Original data from \citet{bentzetal09}; separate determinations for the same object are shown as independent points. The filled (blue) circles refer to NGC 5548; the dashed line is an unweighted least square fit. The filled line is a least-square fit weighted over the uncertainty in $\log t$.}
\label{fig:benz}
 \end{figure}

\section{BLR radius from Reverberation Mapping}
\label{revmap}

The last 25 years \citep[the first recorded attempt is due to ][]{gaskell88} saw the rise of reverberation mapping based estimates of \mbh\  and many reviews can be found on this technique   \citep{peterson93,petersonhorne06,peterson08}.  The cross-correlation function between continuum and emission-line light curves yields a time lag measure $t$\ \citep{gaskellsparke86}. The time lag  due to the light travel time across the broad line emitting region yields an estimate of \rb $= c t$\ \citep{lyutyicherepashchuk72,cherepashchuklyutyi73}.  One can then derive  the black hole mass using FWHM \hbbc\  under  the assumption of virialized motions, using Eq. \ref{eq:mass}. Here we will focus on  the accuracy of such \mbh\ estimations. The measure of \rb\  is based upon several assumptions \citep{gaskellsparke86,krolik01,marzianietal06}. 
\begin{enumerate}
\item The continuum emitting region is much smaller than the line emitting region. This assumption might not be satisfied  since the width of the continuum autocorrelation function  is usually not much less than the measured delay. If the continuum is emitted by the accretion disk, the region emitting the optical continuum might reach the inner BLR \citep[see Table 1 of ][]{gaskell08}. However, the autocorrelation function of the ionizing continuum (usually not observed) might  well be  narrower than the one considered in most reverberation studies \citep[see e.g.,][for a review]{gaskell08}. For \hb, the continuum might be monitored with broad band photometry, or one may measure the specific continuum flux at a suitable wavelength;
\item The observed continuum and the \hi\ ionizing continuum are directly related. The latter assumption  appears to be valid, since the monochromatic luminosity at selected UV wavelengths has been shown to strongly correlate with the BLR radius by the international AGN Watch campaigns (\citealt{claveletal91,petersonetal91,edelsonetal96}; see also \citealt{mclurejarvis02,vestergaard02} and \citealt{gaskell08} for an analysis of NGC 7469, the only object for which a significant time delay between the optical and UV continuum was found). 
\item The light travel time across the BLR is shorter than the dynamical response time $t_\mathrm{dyn}$ (so that BLR structure does not
change over the light travel time). This has been verified by the monitoring campaigns.
\item No measurable dynamical effects due to radiation pressure are present.  This is probably the case for relatively short time lapses ($\le 1$ yr in nearby Seyfert 1 galaxies) i.e., a timescale  shorter than  $t_\mathrm{dyn} \sim $ \rb/ $\Delta v_\mathrm{r} \approx 4 $ yr in the case of  NGC 5548, by far the most extensively monitored object. 

%It is not clear whether there is an actual cloud displacement or whether the effect is due to a light echo crossing a BLR with significant thickness (or whether both effects are at play). 
%In addition, some peculiar structures in the line profiles of sources like OQ 208 might  result directly from radiation pressure
%\citep{marzianietal93}. 

\item There is a well-defined radius that characterizes the BLR. A source  of intrinsic uncertainty involves the likelihood that the BLR is not a thin shell of gas at the estimated \rb\  but a rather thick and stratified structure. This means that the power spectrum of spatial frequencies in the cross correlation function of continuum and line data can be complex.

% -- in fact, as pointed out right above, the BLR radius will likely  fluctuate or "bounce" in response to continuum changes  \citep{mathews93,netzermarziani10}.

\item The line response is linear. The responsivity of the Balmer lines are generally anticorrelated with the incident photon flux \citep{pronikchuvaev72}. Thus, the responsivity varies with distance in the BLR for fixed continuum luminosity and it can change with time as the continuum varies \citep{koristagoad04}. This effect might help isolate a particular region that is truly virial.

\item Line emission is roughly isotropic. This assumption is unlikely to hold for Balmer lines. In this case the angular pattern of emissivity should favor the irradiated face of the line emitting cloud \citep{ferlandetal92}.
\end{enumerate}

More technical problems involve unevenly sampled data, errors in flux calibration, normalization of spectra (different setup, aperture effects), and dilution by stellar continuum \citep{peterson93,horneetal04,bentzetal06,bentzetal09}. This last problem has been found to seriously affect  estimates of the index in the \rb - $L$ relation. According to \citet{krolik01},  effects of a broad radial emissivity distribution (item {3}), an unknown degree of anisotropy (item 7), and  poor sampling in line and continuum monitoring can cause additional {\em  systematic errors} in \mbh\ determination  as large as a factor of 3 or more in either direction.  

With reference to item 4, things appear to be different on a  timescale longer than a couple of years. Looking in  more detail at  the  behavior of NGC 5548, and separating the 13 years of  data points of continuum and \hb\ flux  on a year-by-year basis reveals a distinct correlation between time delay and continuum flux  \citep{petersonetal02}.   The BLR appears to breath  in response to continuum changes   involving a  significant change in \rb\ by a factor $3$\ if we exclude some outlying data points. Even if the line width decreases with increasing \rb\ (as expected for virial motions), an application of Eq. \ref{eq:mass} yields  a somewhat different mass  at each epoch when the continuum luminosity is different --  in principle an absurd result. Using the data for \hb\ of NGC 5548 \citep{petersonetal04}, we find that \mbh\ changes by a factor 4 if the year-by-year data are separately used. If they are averaged together,  $\overline{M_\mathrm{BH}}  \approx (1.3 \pm 0.3) \cdot 10^{7}$ \msol, where the uncertainty is at 1$\sigma$ confidence level\ and could be taken as indicative of the {\em maximum  statistical accuracy} achievable on the virial product \rb $\sigma^{2}$\ using the reverberation method. 

We remark that this degree of accuracy  is not possible  for the wide majority of the sources with \rb\  determination from reverberation data. A major limitation of the reverberation technique involves the need for many spectra of an individual source and spectra spaced across a large enough time interval to capture well-defined continuum fluctuations and subsequent line  responses. This requirement led to a worldwide campaigns that resulted in reverberation measures for \hb\ of more than 60 low $z$ sources \citep{bentzetal09,bentzetal10,denneyetal10}. Only a few sources were observed in  campaigns that allowed two or more independent determinations of \mbh. Differences in derived \mbh\ can be as large, a factor 3 for Mark 817 and a factor 2 for Mark 110 \citep{petersonetal04}. 

There is a tight correlation  between  X-ray variability amplitude and  \mbh\  estimates for sources with \hb\ reverberation data. The intrinsic dispersion in a  best-fit linear relation is   0.2 dex \citep{zhouetal10}. The method based on X-ray variability relies on the assumption that the X-ray variability timescale correlates with source size \citep[e.g., ][]{hayashidaetal98,nikolajuketal04}. The X-ray method has the advantage that it does not rely on the virial assumption. The agreement is very good and the rms scatter quoted above is probably representative of a {\em typical  statistical accuracy} of \mbh\ values derived from the reverberation method. 

A few reverberation measures of \civ\ also exist using IUE  (e.g., \citealt{gaskellsparke86,koratkargaskell91,turlercourvoisier98} and the early International AGN Watch campaigns e.g., \citealt{claveletal91,koristaetal95}) and HST \citep{petersonetal04}.  However, the \civ\ line profile is probably composed of one or more components whose broadening is either non-virial (\citealt{sulenticetal07}; see also \S\ \ref{danger})  or affected by scattering \citep[e.g.,][]{gaskellgoosmann08}. The interpretation of the \civ\ reverberation data is therefore expected to be even more complex than the one of \hb. Unambiguous results may be possible only if velocity resolved reverberation mapping is carried out.

\section{A more widely applicable technique: \mbh\ scaling with luminosity}

\subsection{The \rb\ - $L$\ correlation}

Clearly we need a technique that can provide reasonable estimates for the BLR radius in many hundreds or even thousands of quasars. Virial estimation appears to be the obvious and only way to accomplish this goal but we require a faster way to obtain estimates of \rb\ for large numbers of sources. 
Reverberation radii are rare and extremely rare for high luminosity ($\log L_\mathrm{5100\AA} > 45$) quasars which are the sources for which we are currently able to carry out only single-epoch FWHM \hb\ measurements beyond $z \sim 0.7$. We proceed  to correlate all reverberation values with measures of source luminosity. The resultant  relation \citep{kaspietal00,kaspietal05,bentzetal09}   provides a secondary estimate of \rb. We assume  that the BLR scales with luminosity beyond $\log L \approx 45$\ so that the relation defined for low luminosity quasars can be directly employed for quasars up to 2  dex higher luminosity. It is now standard practice to estimate the BLR radius by assuming  

\begin{equation}
 r_{\rm BLR}\propto (\lambda L_{\lambda})^\mathrm{a}.  
\end{equation}

The exponent in the  correlation between source luminosity $L$\  and \rb\ is  $0.5 \simlt \alpha \simlt$ 0.3, where $a \approx 0.7$ was obtained by \citet{dibai77}.  The exponent has  usually been assumed to be $a$=0.5 -- 0.7, with $a\approx$0.52  now considered the most reliable value \citep{bentzetal09}. Any deviation from this value has quantitative effects that are rather modest. If a restriction is made to the most likely range,  $0.5 \simlt \alpha \simlt$ 0.6, the effect on \mbh\ estimates is $\approx$ 0.3 dex  over three orders of magnitude in luminosity. 

If we employ \hb, $ L_\lambda$\ is the specific luminosity at 5100 \AA\ in units of \ergss \AA$^{-1}$,  the most homogenous sample available to-date yields the relation shown in Fig. \ref{fig:benz}. 

  %However, changing $\alpha$\ implies an $L$-dependent change in mass estimates; therefore the slope of the  luminosity-to-mass relation  is affected as well as the location of points in the \L/M\ vs. \mbh\ diagram \citep{woourry02,marzianietal03b}. 

It is important to stress that the relation  is mainly based on low $z$ ($ \simlt 0.4$) quasars  and that the high (and low) luminosity ends of the correlation are poorly sampled. Considering only  sources in the range $43 \simlt \log \rm L/L_\odot \simlt 45$\  (i.e. where  sources in \citealt{bentzetal09} show uniform sampling), we still  have two orders of magnitude to go in order to reach the most luminous quasars. Efforts  are now underway for   long term monitoring of high-luminosity quasars, although they have yet to produce significant results \citep{bottietal10,kaspietal07,treveseetal07}. Similarly, there is a  program attempting to cover the lower and middle luminosity range of the correlation, the Lick AGN Monitoring Project (LAMP) \citep{greeneetal10a,bentzetal10,denneyetal10}. As it is evident from Fig. \ref{fig:benz} where yearly derivations of \rb\ for  NGC  5548 are shown in blue, there will be some irreducible intrinsic scatter. Using the $L$-\rb\ relation yields a loss of accuracy for \mbh\ estimation; from Fig. \ref{fig:benz} one infers that the scatter is at least a factor 2 at a 2$\sigma$\ confidence level at $\log L \approx 43.5$, although a factor 6 seems  possible at lower luminosity where data points a may also begin to deviate   from linearity. 

\begin{figure}
 \hspace{-0.5cm} \epsfxsize=15cm \epsfysize=15cm 
 \epsfbox{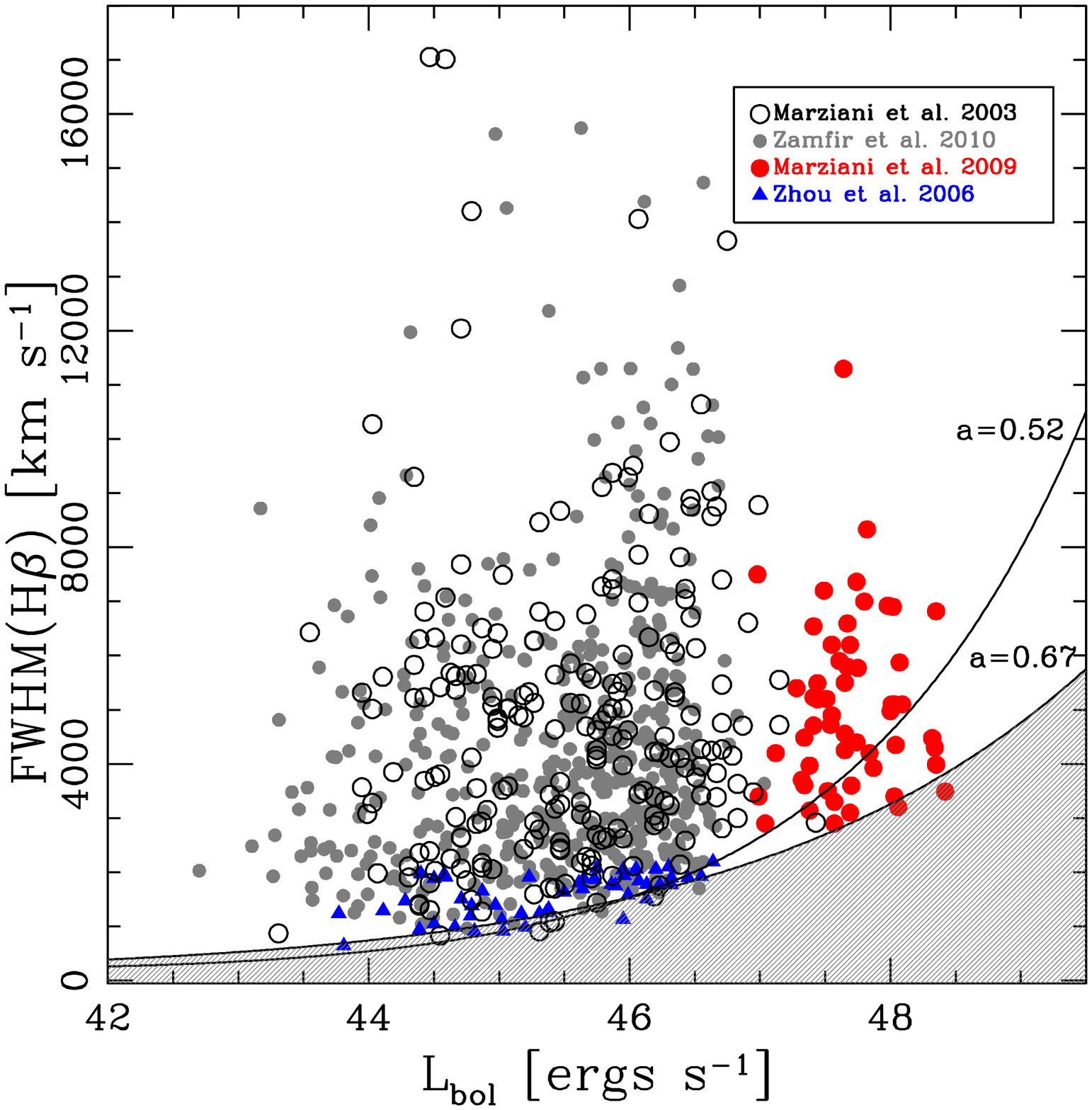}
% \epsfxsize=5cm \epsfysize=5cm
 % \epsfbox{civnc2.eps}
 %\epsfxsize=5cm
 %\epsfysize=5cm 
 %\epsfbox{civnc3.eps} 
 \caption{FWHM as a function of bolometric luminosity  reported for two low-$z$\ samples and the high-luminosity, intermediate $z$ sample of \citet{marzianietal09}. The last sample is based on high s/n  observations of \hb\ with the IR spectrometer ISAAC at VLT  \citep{moorwoodetal98}. The shaded area emphasizes an apparent avoidance zone where no source is expected if quasars do not radiate super-Eddington, and the exponent of the \rb-$L$\ scale relation is $a \approx 0.67$. Given that the relevant trend is for the {\it minimum} FWHM as a function of $L_\mathrm{bol}$, we show the narrowest sources of a large SDSS sample of NLSy1s  \citep{zhouetal06}. These sources also follow the expected {minimum} FWHM dependence on $L_\mathrm{bol}$. Adapted from \citet{marzianietal09}. }\label{fig:fwhm}
 \end{figure}

Fig. \ref{fig:fwhm} shows FWHM  \hb\ as a function of luminosity for several samples, including the high-luminosity sample described in \citet{marzianietal09}. The minimum FWHM  \hb\ apparently increases with luminosity. The plotted lines  refer to an \rb - $L$\ relation with indices $a=0.52$ and $a=0.67$,   assume virial motion and sources at each $L$ radiating at Eddington limit (i.e., the limiting curve are not an effect of any flux-limit condition). In principle, the trend observed can be used to constrain the exponent of the   \rb\ - $L$\ correlation if large samples of high-$L$ quasars are used. The data presented in \citet{marzianietal09} for intermediate $z$ quasars apparently favor a value   larger than $a=0.52$, although the lowest-FWHM points  could be due, for example, to orientation effects  (\S \ref{orient}).

\subsection{\mbh\ - $L$\ relations }

The correlation between \rb\ and luminosity has been redefined for the UV continua luminosity appropriate for using the width of UV lines as a virial broadening estimator. This simply means to compute a best fit between continuum at  3000 \AA\  for \mgii\ (or at 1350 \AA\ for \civ) and the \hb-based \rb\  \citep[e.g., ][]{mclurejarvis02,mcluredunlop04,vestergaardpeterson06,shenetal08,rafieehall11}. 

\citet{vestergaardpeterson06} derived mass scaling relations from the empirical \rb -- $L$\ relation that have been used to compute \mbh\ for large samples of quasars with single-epoch \hb\ or \civ\ observations. \citet{vestergaardpeterson06}  then applied the scaling relations to the objects whose \rb\ is available from reverberation mapping.  They found that the  statistical accuracy of  \mbh\ mass estimates is  low,  $\approx \pm$0.66 and $\approx \pm$0.56 at a $\pm 2  \sigma$ confidence level for \hb\ and \civ\ respectively. Improvements could come from a more refined consideration of the virial broadening estimator for single epoch observations (\S \ref{popab}).

%\citep{wangzhang03}.  
 
\section{Virial broadening and distinguishing quasar populations}
\label{popab}

Single epoch measures of FWHM  are not necessarily the best estimators of the BLR velocity dispersion. The highest precision measure of the virial product involves emission-line width obtained using the cross-correlation function centroid (as opposed to the cross-correlation function peak) for the time delay and  line velocity dispersion $\sigma$ (as opposed to FWHM) for the line width. Obviously one wants to measure the line width in the variable part of the profile. This is the  part most likely to be optically-thick to the Lyman continuum and  therefore responding to continuum changes. \citet{petersonetal04} found that that the rms $\sigma$  minimizes the random component of  errors in reverberation-based \mbh\ measurements (with an absolute minimum around 30\%\ for NGC 5548, as  discussed in \S \ref{revmap}), so that rms $\sigma$ is probably the best choice.

%Line profiles of quasars differ on the basis of the ratio FWHM/$\sigma$.  %By comparing the relative merits of using straight FWHM measures vs. the second moment of the line profile $\sigma$ as a vririal broadening estimator, \citet{collinetal06} confirmed the existence of two quasar populations already noted in \citet{sulenticetal00a,sulenticetal00b}.  

Which measurement is better in general for single epoch spectra, FWHM or $\sigma$? FWHM is very straightforward; however it is sensitive to substructures in the profile (e.g., double peaked profiles), and presence of inflections that can yield unstable values depending on continuum placement. On the converse, $\sigma$\ has the significant disadvantage of diverging in profiles with prominent line wings (and to be $\rightarrow\infty$\ for Lorentzian profiles). Especially in the case of low s/n,  $\sigma$\ values are affected by large errors. 

One must at least  consider that the ratio FWHM/$\sigma$\ changes with line width \citep{collinetal06} and avoid mixing all sources together in large samples \citep[e.g., ][]{nagaoetal06}. The change occurs when radiative and gravitational forces appear to balance (Eq. \ref{eq:radgrav}): \lledd $\approx 0.15 \pm 0.05$ \citep{marzianietal03b} for \nc = $10^{23}$ cm$^{-2}$, probably an appropriate average value of column density.  %\citep{marconietal08}. 

\begin{figure}
\hspace{-0.5cm} 
 \epsfxsize=7cm 
 \epsfbox{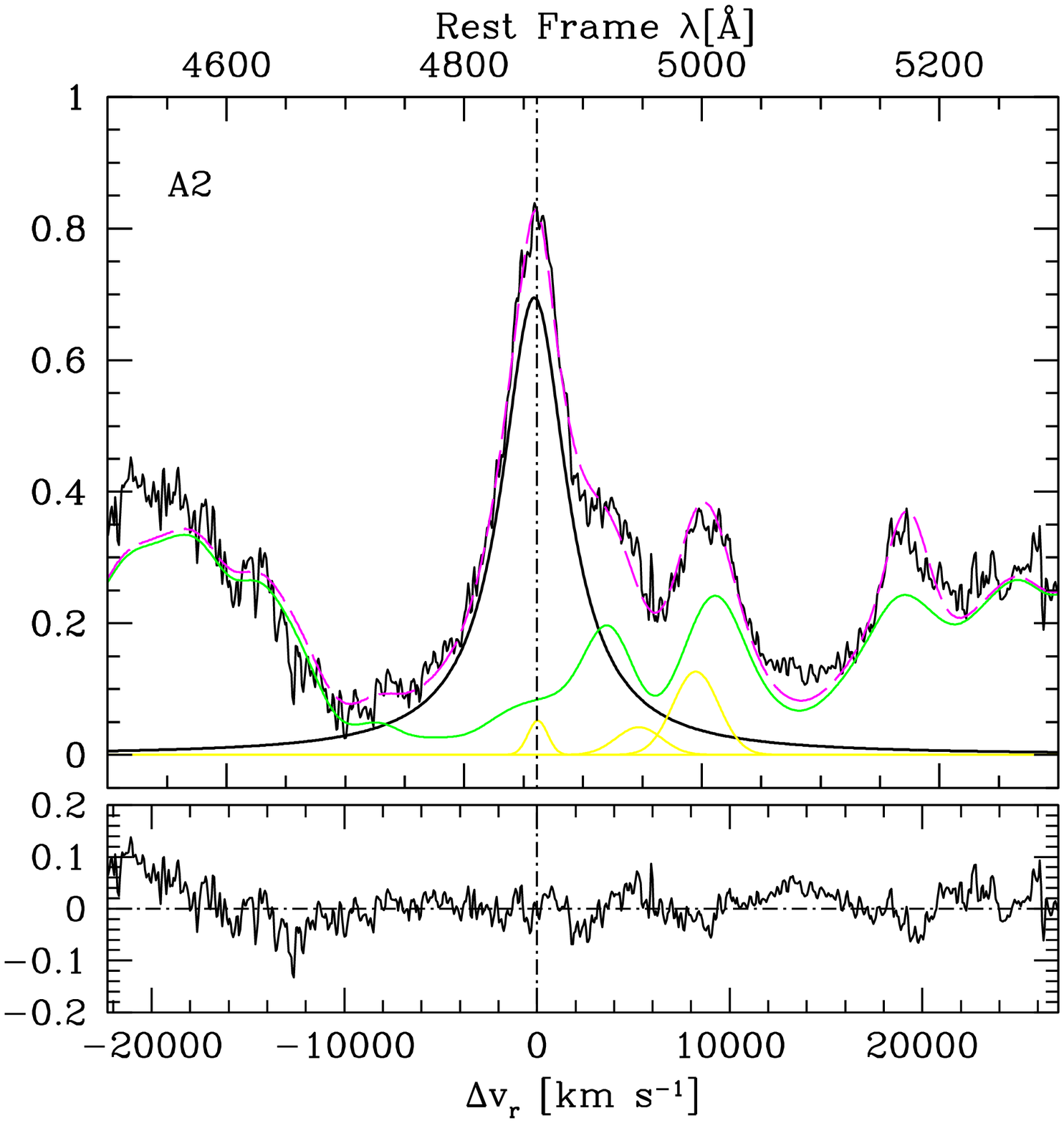} 
 \epsfxsize=7cm 
 \epsfbox{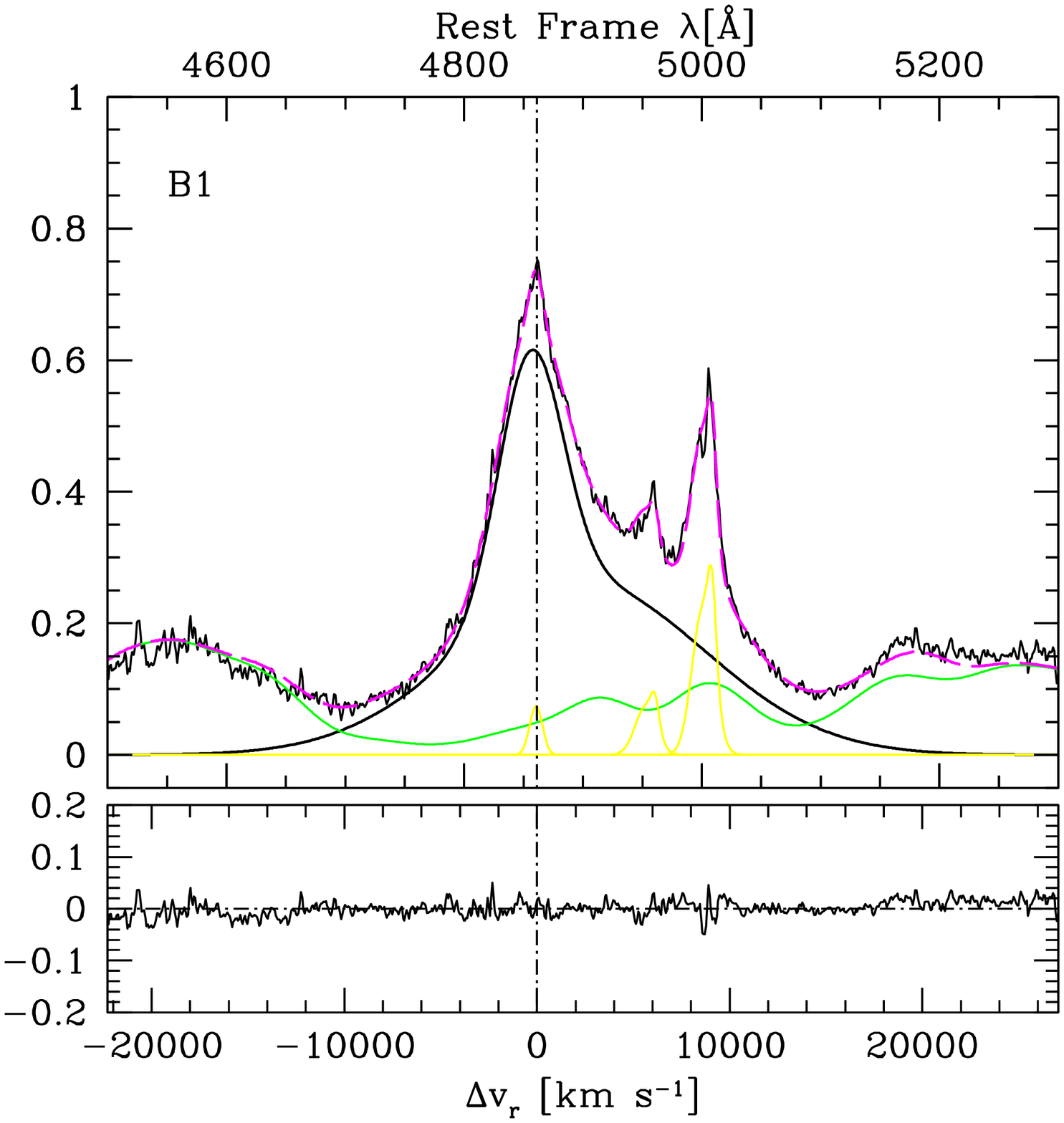} 
 \caption{ \hb\ profiles in spectral types A2 and B1 that are typical of Pop. A and B respectively. The displayed median spectra have been computed from a large sample of SDSS high s/n, low-$z$\ quasars (Sulentic et al., in preparation). }
 \label{fig:ab}
\end{figure} 

\subsection{\hb\ in Population A Sources}

%Sources with Pop. A \hb\ profiles  yield results that are quite robust to the detailed technique applied. 
Pop. A profiles (FWHM \hb $<$  4000 \kms) are  well fit by a single component Lorentz function \citep[Gaussians yield larger residuals; ][]{veroncettyetal01,sulenticetal02,marzianietal03b} and tend to show lower amplitude variations and less  change in profile shape when continuum varies (compared to Pop.  B(roader) sources; \citealt{klimeketal04,aietal10}). One can describe them as Narrow Line Seyfert 1-like  sources. The biggest challenge for extracting an accurate FWHM measures from Pop.  A profiles involves careful subtraction of the usually strong \feii\  emission that afflicts and broadens the red side of the profile. Line profile shifts or asymmetries are usually small.    

\subsection{\hb\ in Population B Sources} 

The typical  Pop. B H$\beta$ line  (FWHM \hb\ $>$ 4000 \kms) is more complex and requires two Gaussian component  models to adequately describe the profile (Fig. \ref{fig:ab}). The narrower, broad component is likely the most reliable virial estimator while the VBC component showing a redshift of 1000 - 2000 \kms\ and FWHM~~10000 \kms\ is unlikely to be virial. The narrower component then is viewed as the classical BLR component which is the only one seen in Pop. A sources. As for Pop. A, usually  shifts are small although one must consider that a fraction of Pop. B sources show shifts that are significant relative to the line width; \citealt{zamfiretal10}). It apparently reverberates more strongly than the VBC \citep[e.g., ][]{sulenticetal00c,koristagoad04}. The VBC can affect both estimates of \rb\ and virial velocity for almost half of quasars. It must be taken into account during reverberation studies or it will slow down and/or blur the CCF results; it must be considered in measures of FWHM \hb\ or the virial velocity can be seriously overestimated (factors of 2 or 3). The relative strengths of the two \hb\ components appears to depend on source luminosity \citep{marzianietal09,meadowsetal11}  with the VBC becoming more dominant in high luminosity quasars. This is obviously important for studies of \rb\ and \mbh\ as a function of $L$ and $z$. As spectral s/n decreases the two components are unlikely to be recognizable with the result that FWHM   \hb\ will be systematically overestimated.

\section{Mass estimates for large samples}

We must  examine all strong broad lines as potential virial estimators. In the following, this  review focuses on attempts to estimate \mbh\  for large numbers of quasars over as wide as possible range of redshift and source luminosity. Opinions will vary about how far we have succeeded in this quest.    

Whether Pop. A or B, the number of sources with reliable FWHM H$\beta$ measures becomes increasingly small above $z \approx$ 0.7. One is left with three options if one wishes to study sources at higher redshift: (1) follow H$\beta$ into the infrared, or (2) adopt other broad lines (e.g. \mgii\ or \civ) as surrogate virial estimators. If one wishes to study lower luminosity sources at low redshift then option (3) involves using databases like SDSS where contextual binning of many source spectra can yield high s/n profiles \citep[see e.g. ][]{zamfiretal10,meadowsetal11}. The availability of infrared spectrographs at 4m+ class telescopes makes option (1) possible for tens or even hundreds of sources. By ``possible'' we mean that it becomes possible to obtain spectra with s/n ($\sim$ 20 in continuum near H$\beta$) comparable to lower redshift optical measures for the brightest individual SDSS (or other) spectra. Using  ESO VLT and  IR spectrometers like ISAAC  one can presently obtain such spectra for sources at the bright tip of the quasar luminosity function up to $z \simlt 3.8$\ with exposure times of $\sim$40 minutes. The availability of IR windows determines the source redshift distribution using H$\beta$ but reasonable coverage over the range $z = 1 - 3.8$\ is possible. The advantage of this approach is uniformity, same reduction procedure and same virial estimator. 

The SDSS database is a seductive  source of data for more than 100000 quasars offering the possibility to derive instant \mbh\  estimates. Beyond the brightest $\approx$ 500 sources this is an illusion as the decline in s/n makes reliable FWHM estimates increasingly difficult. %After all the spectra were obtained with a 2.5m telescope. 
If all quasar spectra were basically the same then one could proceed by fitting profile models, determined for the brightest sources, to large numbers of noisy spectra. We now know that spectra of quasars are less similar than are stellar spectra along the main sequence of the H-R Diagram. Dissimilarities include: 
\begin{enumerate}
\item  internal broad (and narrow) line shifts, \item the line profiles at a fixed redshift or source luminosity show large differences as detailed earlier, \item \feii\  contamination varies widely and directly affects the FWHM measure. 
\end{enumerate}
Quite aside from arguments about the validity of any specific line as a virial estimator, the SDSS database of spectra (all but 100s) can be used but only if the data are binned into some well defined context (for example the spectral type defined by \citealt{sulenticetal02}), or some innovative approach is applied \citep[e.g., ][]{rafieehall11}. 

\subsection{Mass estimation using \mgii}

\citet{mcluredunlop04} computed virial black hole mass estimates  for a sample of more than 10000 quasars in the redshift interval $ 0.1 < z < 2.1$ drawn from the SDSS Quasar Catalogue  of \citet{schneideretal03}. The  \hb\ line was used up to $z\approx 0.7$, and  \mgii\  beyond. The change in virial estimator did not introduce any obvious discontinuity.   Also no evidence of large masses exceeding the \mbh\ of locally detected dormant black holes were found. This behavior is not surprising since \hb\ and \mgii\ are both LILs thought to be produced mostly in the same emitting region. There is evidence that the \mgii\ FWHM distribution might be narrower than the one of \hb\,   \citep{labitaetal09}, and  that individual \mgii\ FWHM values are systematically narrower than \hb\, \citep{wangetal09}.   If so (and if the presence of narrow absorptions on the blue side of \mgii\ are not creating a spurious effect), \mgii\ might be more appropriate because  its single epoch profile may be a better approximation of the reverberating part of \hb\ isolated on the rms spectrum, as also suggested for \feii\ \citep{sulenticetal06}.  

%\citep{vestergardpeterson06} 
 
\subsection{(Dangerous) mass estimation using \civ}
\label{danger} 
 
\begin{figure}
 \hspace{-0.5cm} 
 \epsfxsize=7cm 
 \epsfbox{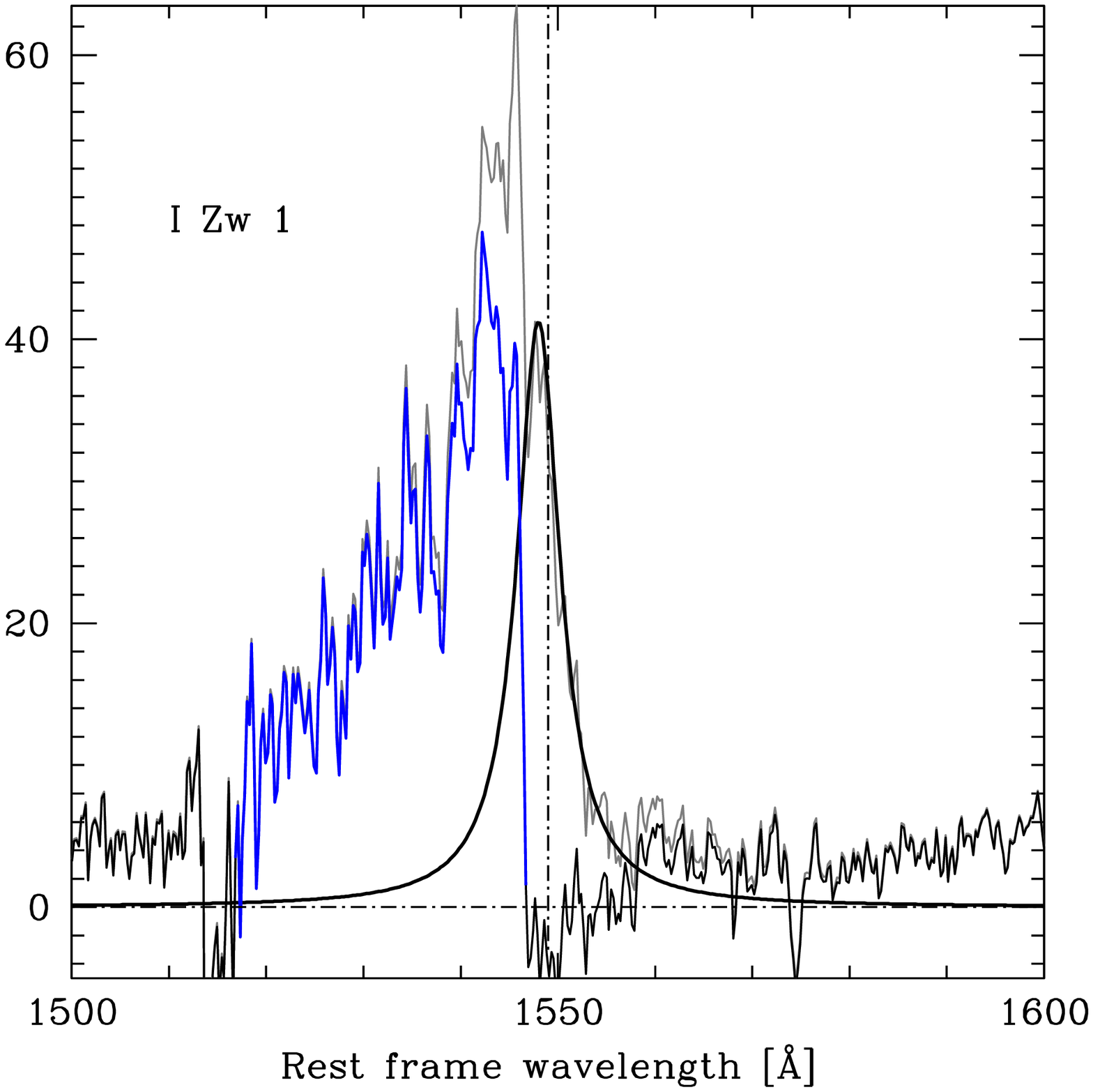}
 \epsfxsize=7cm 
 \epsfbox{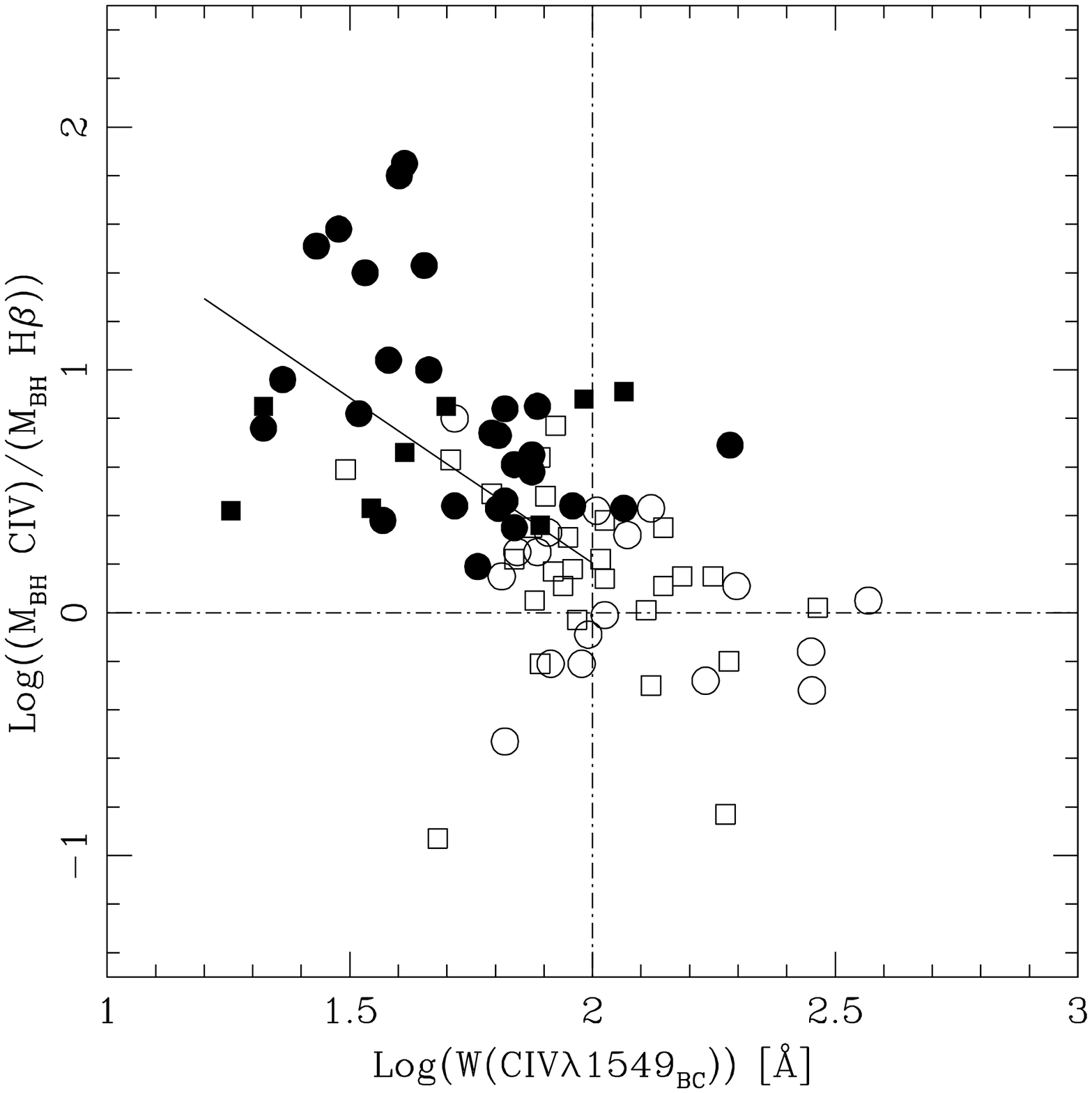}
  \caption{Left: \civ\ and \hb\ profile of the prototype Pop. A source I Zw 1. The grey line represents the original profile, the black one the scaled \hb\ profile and the blue solid line the residual after subtraction of the scaled \hb\ profile.  The ratio between \mbh\ computed using FWHM \civ\ and FWHM \hb\ as virial broadening estimators as a function of W(\civ). Filled circles: Pop. A radio-quiet sources; filled squares: Pop. A radio-quiet; open circles: Pop. B radio-quiet; open squares: Pop. B radio-loud.  Adapted from \citet{sulenticetal07}. }
 \label{fig:civhb}
 \end{figure} 
  
Much more controversial has been the use of \civ\ as a virial broadening estimator \citep{baskinlaor05b,netzeretal07}. 
 
The most  serious effect is related to the strong blueshift with respect to the rest frame frequently  observed in   \civ\  \citep{gaskell82,richardsetal11}.   This suggests that either the high-ionization line emitting gas is outflowing in a wind \citep{gaskell82,marzianietal96} in at least a significant fraction of quasars and hence not in virial equilibrium, or that there is some other broadening mechanism, such as electron scattering or Rayleigh scattering \citep{gaskellgoosmann08}. In either case \civ\ is not a reliable mass indicator.  
This  fraction may even increase with redshift and luminosity, judging from the observed \civ\ profile in a high $z$ quasar sample \cite{bartheletal90}.  The effect is very pronounced if we consider a prototypical source believed to accrete close to the Eddingon limit, I Zw 1 (Fig. \ref{fig:civhb}). The \civ\ profile can be modeled as an the unshifted almost symmetric \hb\ profile plus a prominent blueshifted component. The broadening, in this case, is dominated by a shifted non-virial component. Since \mbh\ $\propto (\delta v)^{2}$, the effect can reach a factor 10 on \mbh, as shown in the right panel of Fig. \ref{fig:civhb}.  For Pop. B sources, there is no obvious systematic effect, although the \mbh\ derived from \civ\ is sometime much lower than the one derived from \hb, probably because of  significant \civ\ narrow emission (\civnc).
  
 The very existence of \civnc\  has been a contentious issue. Radio-loud sources often show prominent \civ\ cores of width $\sim$ 2000 \kms, but such emission has been ascribed to \civbc\ \citep{willsetal93} and to an  intermediate line region \citep{brothertonetal94a}. We have interpreted this feature as part of the NLR, and recent detection of narrow \civ\ in high $z$ type-2 sources supports this interpretation \citep[e. g., ][]{mainierietal05,severgninietal06}.
  Provided that a smooth density gradient exist (i.e.,
 that there is a significant amount of gas at $n_\mathrm{e}$\ $\sim$10$^6$
 \cm3, we expect \civ\ emission to be strong (\civ\ emissivity is
 $\propto n_{\rm e}^2$ while \oiiiopt\ should become collisionally
 quenched). If velocity dispersion increases with decreasing
 distance from the central continuum source, as expected if there
 is at least rough virial equilibrium, then it is somewhat natural
 to expect FWHM\-(\civnc) $>$ FWHM\-(\oiiiopt) \citep{derobertisosterbrock86,appenzellerostreicher88,sulenticmarziani99}.  
 
 %There is plenty of evidence
 %that at least some AGN with strong narrow lines have a strong
 %\civnc. The visual impression (already confirmed by \citep{marzianietal96}) of a
 %connection between W(\oiiiopt) and the W(\civ) of the narrow core
 %is expressed through a loose correlation between W(\oiiiopt) and
 %W(\civnc) which we found considering {\em only} the cases of a
 %least ambiguous inflection. Since \civnc\ is prominent in Pop. B sources but can be
 %even absent in Pop. A, the wind that gives rise to the
 %blue-shifted \civbc\ component may serve as a filter to the FUV
 %ionizing radiation from the center that would otherwise reach the
 %outer self-gravitating parts of the disk, where \civnc\ may
 %originate \citep{leighly04}. 
 
 A comparison of FWHM \hbbc\ and \civ\ line  shift measures with those of other workers \citep{warneretal04,baskinlaor05b}   shows that significant and systematic differences exist between the  measures.  FWHM \civbc\ measures  begin to deviate towards smaller values  at FWHM \civ\ $\sim$ 4000 \kms\ where we believe that
 the NLR becomes important (Pop. B; \citealt{bachevetal04}).   However, considering the strong  Baldwin effect that is observed in \oiii\ (mainly due to the increasing rarity of \oiii-strong objects at high luminosity \citealt{netzeretal04,kovacevicetal10}; Sulentic et al., in preparation), the relevance of   \civnc\  could be small in high $z$ quasars, for which \civ\ becomes the only line available for estimating the virial broadening from optical observations. 
 
Concluding, the FWHM of \civ\ is a poor virial broadening estimator  subject to systematic biases. Alternatively the intermediate ionization lines (especially \siiii\ and \aliii) could  in principle be used even if  blended. Intermediate-ionization lines like \siiii\ and \aliii\ show negligible blueshifted, non virial emission, weak VBC and  a profile roughly consistent with the broad component  of   \hb\ \citep[][and references therein]{marzianietal10}. 

 \section{Photoionization Methods}

Historically, the first attempt to derive \mbh\ from photoionization conditions is due to \citet{dibai77}, as mentioned in \S \ref{intro}. The underlying assumption was of course that all AGN were in similar physical conditions, so that the luminosity of \hb\ could be related to the BLR size assuming an identical value of the line emissivity. The application of Eq. \ref{eq:dibai} needs an estimate of the filling factor. If spherical symmetry is assumed,  $f_\mathrm{f}$ can be easily computed from the covering factor $f_\mathrm{c}$ that is observationally constrained \citep{wandelyahil85}. 

The physical conditions of photoionized gas can be described by  hydrogen numeric density $n_\mathrm{H}$\ or electron density, hydrogen column density $N_\mathrm{c}$\, metallicity,  shape of the ionizing continuum, and the ionization parameter $U$.  The latter represents the dimensionless ratio of the number of ionizing  photons and the total hydrogen density, ionized and neutral.  If we know {\em the product of } $n_\mathrm{H}$\ (or electron density \ne) and $U$, we  can estimate \rb\ from Eq. \ref{eq:u}. $U$ can be written in terms of \rb\ yielding an estimate of the BLR independent of direct reverberation estimates:

\begin{equation}
r_{\rm BLR} =  \left( \frac{\int_{\nu_0}^{+\infty} \frac{L_\nu}{h\nu} d\nu}{ 4 \pi U n_{\rm H} c} \right)^\frac{1}{2}
\propto \frac{1}{(U n_\mathrm{H})^{\frac{1}{2}}} {Q(H)}^{\frac{1}{2}}
\end{equation}

The dependence of $U$\ on $r_{\mathrm{BLR}}$ was used by \citet{padovanirafanelli88} to compute central black hole masses assuming a plausible  average  value of the product $n_\mathrm{e}  U$. \citet{padovani89} derived an  average  value $\overline{U \cdot n_\mathrm{e}  }  \approx 10^{9.8}$\ from several sources for which $r_{\mathrm{BLR}}$\ was available from reverberation mapping and for which the number of ionizing photons could be measured from multiwavelength observations. The average $n_\mathrm{e}  U$\ value was then used to compute black hole masses for a much larger sample of Seyfert 1 galaxies and  low-$z$\ quasars \citep{padovanirafanelli88,padovanietal90}. Multifrequency data were used to define the shape of the ionizing continuum for each individual source. \citet{wandeletal99} compared photoionization method results with those obtained via reverberation mapping and found a very good correlation between the two mass estimates.

Recently, the photoionization method of \citet{dibai84}  based on   \hb\ luminosity was reconsidered \citep{bochkarevgaskell09}.  The original Dibai data yield   very good agreement with   \mbh\ estimates from reverberation-mapped sources.

A different photoionization method could be applicable to high redshift quasars. Emission lines originating from forbidden or semi-forbidden transitions become collisionally quenched above the critical density and, hence, weaker than lines for which collisional effects are still negligible. The Al{\sc iii}$\lambda$1860 /Si{\sc iii}]$\lambda$1892\ ratio is well suited for exploring the density range $10^{11} - 10^{13} $ cm$^{-3}$\ which corresponds to the densest, low ionization emitting regions likely associated with the production of Fe{\sc ii}.   The  ratios  Si{\sc ii}]$\lambda$1814 /Si{\sc iii}]$\lambda$1892 and Si{\sc iv}$\lambda$1397/ Si{\sc iii}]$\lambda$1892  are independent of metallicity and sensitive to ionization. Conversely the ratio C{\sc iv} $\lambda$1549/Si{\sc iv} $\lambda$1397 is mainly sensitive to metallicity.  

We computed a multidimensional grid of {\sc CLOUDY} \citep{ferlandetal98} simulations \citep[see also ][]{koristaetal97} in order to derive $U$\ and $n_\mathrm{H}$.   Computation of constant value contours  in the theoretical $U$ vs. $n_\mathrm{H}$\ plane of {\sc CLOUDY} simulations shows convergence towards a low ionization plus high density range. Application of this method to the  NLSy1 I Zw 1  and NLSy1-like sources at high redshift provides not only an independent estimate of $nU$ but also of \nh, $U$, and metallicity \citep{negreteetal10}. \rb\ values  obtained with the photoionization method can be compared to the ones obtained with  sources  reverberation mapping. The two sets of values agree better than the  values estimated using the \rb - $L$ correlation -- and this employing the very sample used to define the \rb - $L$ correlation \citep{negrete11}. 

The large uncertainty associated with the scaling relation that employs \civ\ (present even if the \civ\ virial component can be properly isolated)   makes preferable a one-by-one \mbh\  determination based on physical properties of an emitting region that remains self-similar. The particular value of the photoionization method is that it can be applied over the full redshift range where quasars are observed.

%Further refinements may be coming if the behavior of $U$\ along the 4D Eigenvector 1 sequence (Sulentic et al. 2000ab) is better understood and if \feii\ can be used as a diagnostic \citep{zamanovmarziani02,verneretal04}. 

 \vspace{10cm}
 \medskip

 %Once the stringent (and perhaps unphysical) distances from the
 %central continuum sources set by gravitational redshift are
 %relaxed, we note that our VBLR model can explain \hb\ VBLR
 %luminosity with reasonable shell radii.

 \section{Black Hole Mass and Host Galaxy Mass}
 \label{bulge}
 
The  correlation of  nuclear black hole mass with stellar bulge velocity dispersion  in nearby galaxies  \citep{ferraresemerritt00,gebhardtetal00} suggests a constant \mbh/\mbulge\ ratio. This result has been extensively used (or should we better say abused?) to derive quasar \mbh\ from the ratio  \mbh/\mbulge\ or from a proxy of the stellar velocity dispersion. 

\citet{ferraresemerritt00}  relied on maser optical emission line measurements. \citet{gebhardtetal00} and \citet{merrittferrarese01} added masses obtained from the so-called stellar dynamics method \citep[e.g.][]{kormendy93,magorrianetal98}. Broadly speaking, these \mbh\ determinations have the advantage that the line emitting region is partly resolved, so that a detailed analysis of geometry and orientation effects can be carried out. It is still debated whether stellar dynamics \mbh\ determinations might introduce  a bias in the measured \mbh\ \citep{gultekinetal09}. However, it is more relevant in the the quasar context to assess   whether  a significant \mbh -- \mbulge\ correlation  exists at all, or whether it results from a bias due to the difficulty of detecting a black hole whose sphere of influence is much smaller than the telescope resolution \citep{gultekinetal11}.  In other words, the correlation may refer only to the maximum \mbh\ for each \mbulge.  The most likely condition is that the bias is  mass-dependent, and most severe at low \mbh\ (i.e., higher mass black holes are more easily detected). The true (unbiased) aspect of the relation could be tight at large \mbh\ but with increasing dispersion toward lower masses.  

%While probably valid for large masses, the relation between \mbh\ and  $\sigma_\star$\ should be taken with special care at the low \mbh\ end.% 

An additional complication is that the relation between  \mbh\ and  $\sigma_\star$\ appears to be different for barred and non-barred hosts \citep[][and references therein]{grahametal11}. Intrinsic dispersion at low \mbh\ and  dependence on morphology could overshadow several conflicting claims e.g., a nonlinear relation  between the \mbh\ and the bulge mass for PG quasars \citep{laor01}:  \mbh\  $\propto M_{\rm bulge}^{1.53 \pm 0.14}$.    Another case is the claim that NLSy1s, often   host  in dwarfish galaxies \citep{krongoldetal01} and  barred spirals \citep{crenshawetal03,ohtaetal07} possess under-massive black holes   \citep[e.g.,][]{mathuretal01,chaoetal08}.
 
A similar difficulty might apply also to the correlation between black hole mass and bulge luminosity  \citep{kormendyrichstone95}: the dispersion of data at low-$L_\mathrm{bulge}$\ could make any correlation ill-defined \citep{gaskellkormendy09,gaskell10}. The $L_\mathrm{bulge}$ -- \mbh\ relation suffers from the additional problem that Seyfert  galaxy hosts are brighter than normal galaxies for a given value of their velocity dispersion, perhaps as a result of younger  stellar populations \citep{nelsonetal04}, even if the use of IR luminosity can greatly reduce the scatter \citep{marconihunt03}.
  
Radio-quiet quasars seem to  follow the established \mbh-$\sigma_\star$ relation up to $z \approx$0.5, with a modest evolution  in the redshift range $0.5 \simlt z \simlt 1$  \citep{shieldsetal03,salvianderetal07}. However, the \mbh/\mbulge\ ratio estimated  at low $z$\ should not be mistaken for  an universal constant.
Studies at high redshift find \mbh\ and \mbulge\ values that  indicate overmassive black holes at high $z$ \citep[][and references therein]{targettetal11}. Evolution in the  \mbh/\mbulge\ ratio casts doubts on the physical implications of the relation at low-$z$. 

%t seems that luminous (--24 $>M_{\rm V} >$  --28) quasars (both radio-loud and radio-quiet) are hosted in galaxies which are spheroidal or, at least, possess large bulges \citep{floydetal04}.  

%NLSy1s  seem to be often host  in dwarfish galaxies \citep{krongoldetal01} with a large fraction in barred spirals \citep{crenshawetal03,ohtaetal07}. The issue is still  debated  with some authors suggesting that NLSy1s are obeying the same scaling relation \citep{komossaxu07}, while others suggesting that NLSy1s have undermassive black holes \citep[e.g.,][]{mathuretal01,chaoetal08} perhaps because of different hosts. 

\section{Using \oiii\ as a proxy for stellar velocity dispersion}
 
%The \mbh\ -- bulge mass relation has been  used to derive  $f$   for the reverberation mapped sample. The additional issue here is that, 
If the  \mbh/\mbulge\  ratio  can be derived with good accuracy, then  \mbh\  could be computed   using a proxy for  $\sigma_{\star}$.   There are two major difficulties with using FWHM  \oiii\ as a   proxy for  $\sigma_{\star}$. While  a \mbh\  -- FWHM \oiii\  correlation exists \citep{nelson00},  the scatter is large.  The derived values can be  considerably higher than those calculated using FWHM \hbbc\   \citep{marzianietal03b}. On the one hand,  the narrowest profiles of \oiii\ appear to be sub-virial in the gravitational potential of the host bulge, probably  because the line emitting gas is constrained in non-spherical geometries (see the discussion in \citealt{gaskell09c} and references therein).  On the other hand, low-$W$(\oiii) sources can show FWHM \oiii\ $\simgt$
 FWHM(\hb), seriously challenging the  assumption of gravitational dominance by the bulge potential.   Blueshifted  \oiii\ profiles are observed and likely arise in outflowing gas \citep{zamanovetal02}, possibly associated with a disc wind. The distribution of \oiii\ velocity shifts with respect to \oii\ indicates that the majority of sources are significantly blue-shifted \citep{huetal08a}. The narrow line region (NLR) in blueshifted sources is   not likely to be dynamically  related to the host-galaxy stellar bulge. This is even more true for a minority of sources with   $\Delta v_\mathrm{r} \simlt -300$ \kms\ (the so-called blue-outliers of \citealt{zamanovetal02}) whose NLR may be very compact. This points to a  limiting W(\oiii) $\approx 20$ \AA below which FWHM(\oiii) emission  probably ceases to be a useful mass estimator. Only large W(\oiii) radio-quiet Pop. B sources   may have very extended NLRs  whose  motions are dominated by the stellar  bulge. Very appropriately, the width of the \sii\ or \oii\ doublet has been proposed as an alternative to \oiiiopt\ width \citep{komossaxu07,salvianderetal07}.
 
% Nelson \& Whittle (1996) have compared [O iii] line widths and stellar velocity dispersions in AGNs, generally finding good agreement. For the quantity log???1Ú2O iii??=?????, they find a mean 0:00 ?? 0:01 and a dispersion ?? 1Ú4 0:20, supporting the idea that the motions of the narrow-line region (NLR) gas are largely determined by the gravitational potential of the host galaxy. This is reinforced by the analysis by Nelson (2000), who essentially shows that ??[O iii] and MBH obey equation (1) for AGNs with echo values of MBH. These results support the use of ??[O iii] as a surrogate for ??*. A cau- tion, however, is that [O iii] profiles often have substantial asymmetry and non-Gaussian profiles, possibly resulting from outflow combined with extinction of the far side of the NLR (e.g., Wilson \& Heckman 1985; Nelson \& Whittle 1995).
 
 \section{Masses from Double Peakers}
 
 A small fraction of AGN exhibit very broad and double peaked LILs. The H$\alpha$ BC profile is strikingly peculiar (see  \citealt{sulenticetal00a,eracleoushalpern03,stratevaetal03} for examples).
 Prototype double-peakers  Arp 102B, 3C 390.3, and 3C 332 have been  monitored for more than 20 years  \citep{lewiseracleous06,gezarietal07}. A common property of double peakers  
 involves variability of the profile shape on timescales of months to
 years. This slow systematic variability of the line profile shows similar
timescales to dynamical changes expected in an accretion disk
 \citep{eracleoushalpern03,newmanetal97}.  Five
 year monitoring  of  \hb\ \citep{shapovalovaetal01}  supports rejection of the binary BH hypothesis \citep{gaskell83} for 3C390.3, on the
 basis of the masses required \citep{halpernfilippenko88}. Hot spots, spiral waves and
 elliptical accretion disks have also been suggested  
 \citep{storchibergmannetal03,lewiseracleous06}. 
 
If one can detect changes in the line profile induced by the hot spot, it is in principle possible to
derive the period of the hot spot and hence to estimate \rb\ and \mbh\ in physical units. In the case of Arp 102B one obtains a mass  $\approx (2.2 \pm 0.7)  \cdot 10^8$ \msol\ for the black hole, and  a distance of 4.8$\cdot  10^{-3}$ pc ($\approx 500$ \rg) for the hot spot \citep{newmanetal97}. Derivation of \rb\ in physical units removes the degeneracy introduced by the Keplerian velocity field  (i.e., the velocity scales as (\mbh\  / $r$)$^{-0.5}$, and disk model profiles yield a distance normalized to \rg).  Unfortunately,  gravitational and transverse redshifts show the same dependence  $\propto$(\mbh\  / $r$)$^{-0.5}$ at first order. A gravitational redshift  will not help us solve for \mbh\ and $r$ in physical units;  detection of a radial velocity change with time will do it. Monitoring has been  pursued for sources showing single-peaked line profiles under the assumption that  the disk contribution is masked by emission from a wind or hot spots \citep{bonetal09a,jovanovic10}. This approach provided a new estimate of the black hole mass for NGC 4151 (Bon et al., in preparation).     
 
 \begin{figure}[ht!]
 \epsfxsize=14cm
 \epsfbox{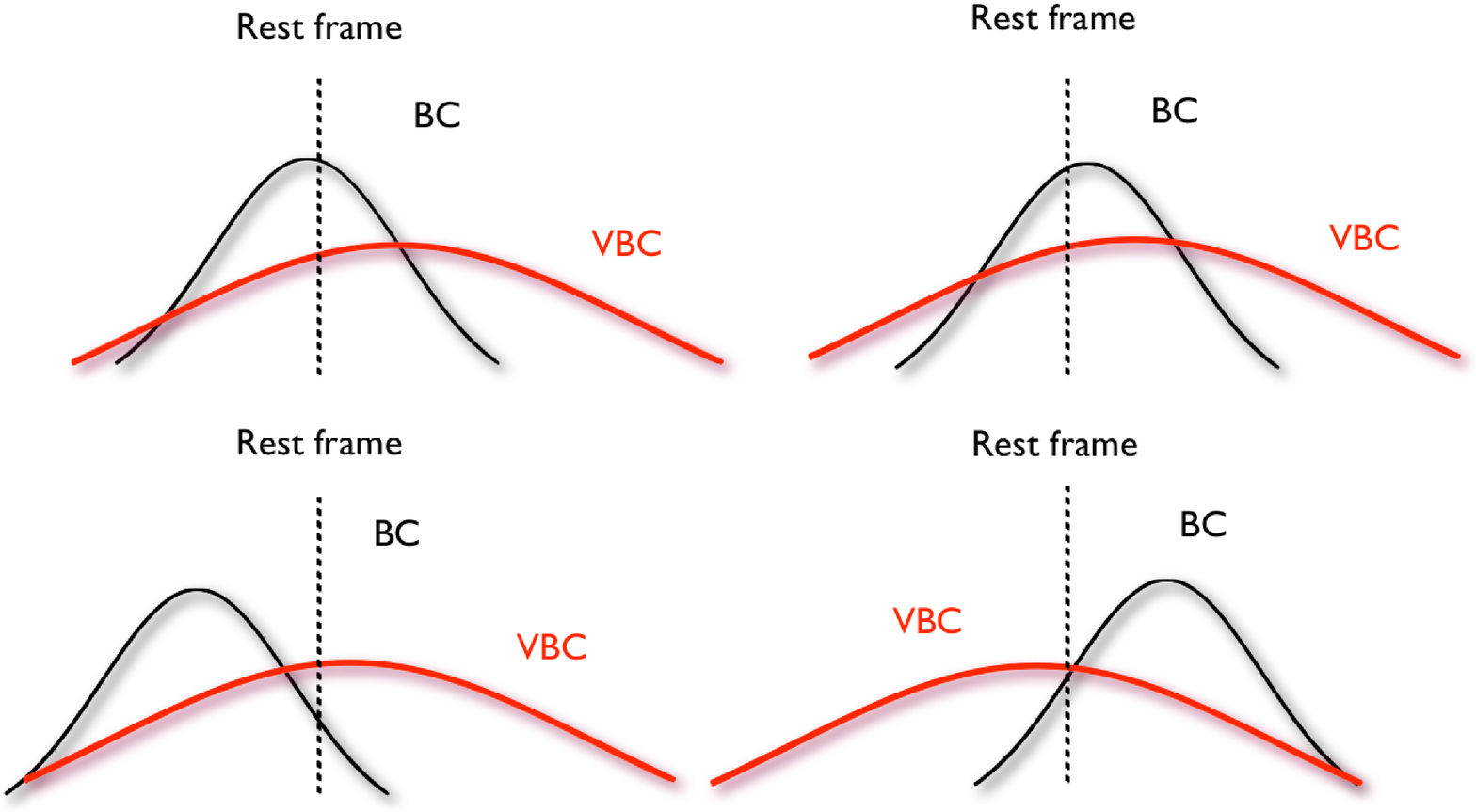}
 \caption[]{Sketch illustrating  2 examples of most-frequently observed H$\beta$ profiles  in Pop. B sources (upper half) and the profiles that might 
be signatures of binary black holes. The SDSS 153636.221+044127.0 \hb\  line profile \citep{borosonlauer09} corresponds to the lower-left case.}\label{fig:bbh} 
 \end{figure}
 
  \section{A binary black hole revival}
 
A binary black hole model was originally proposed to explain sources showing 
double-peaked line profiles like Arp 102B  \citep{gaskell83}.  Lower limits for plausible orbital periods based on the absence of peak radial velocity  changes would require supermassive binary black holes   with total masses in excess of 10$^{10}$ \msol\ \citep{halpernfilippenko88,shapovalovaetal01}. Such large  \mbh\ values are difficult to reconcile with the maximum \mbh\ values observed in the local Universe \citep{eracleousetal97} and with the  \mbh\ -- bulge mass relation.  

The search for supermassive binaries with sub-parsec separation has undergone a revival \citep[][]{eracleousetal11,tsalmantzaetal11} since such binary black holes are expected  in scenarios involving black hole and bulge coevolution \citep{begelmanetal80,volonterietal09}.  A candidate involves the quasar SDSS J153636.221+044127.0 that shows line features  possibly indicating two broad-line systems separated in radial velocity by 3500 \kms, and a  system of unresolved absorption lines at intermediate radial velocity \citep{borosonlauer09}. A  disk model with off-axis illumination has also been proposed \citep{gaskell10b,gaskell10a} for this source.
 
It is difficult to use a single source with almost unique\footnote{A twin of SDSS J153636.221+044127.0 is shown in Fig. 1 of \citet{gaskell83}.} line properties  to argue for a phenomenon (i.e. binary BH) expected to be common among quasars.  Peculiar shifts of the kind identified by \citet{borosonlauer09} are restricted to Pop. B sources and the most  frequently observed profile showing a redward asymmetry and a peak displacement to the red is clearly unsuitable as a binary BLR signature where the NLR indicates the rest frame (Fig. \ref{fig:bbh}). If we interpret the \hb\ BC and VBC as two different BLRs associated with the binary, and we assume that the line emitting gas forms two bound systems, then the largest shifts should be observed in the BC, the contrary of what is most frequently observed (the VBC is redshifted by $\simgt$ 1000 \kms).  Other systems have been selected as possible candidates, notably the ones showing opposite peak and line base displacement in radial velocity \citep{shenloeb10}, but altogether they account for no more than a few percent of quasars.

\begin{figure}[ht!]
 \epsfxsize=15cm
 \epsfbox{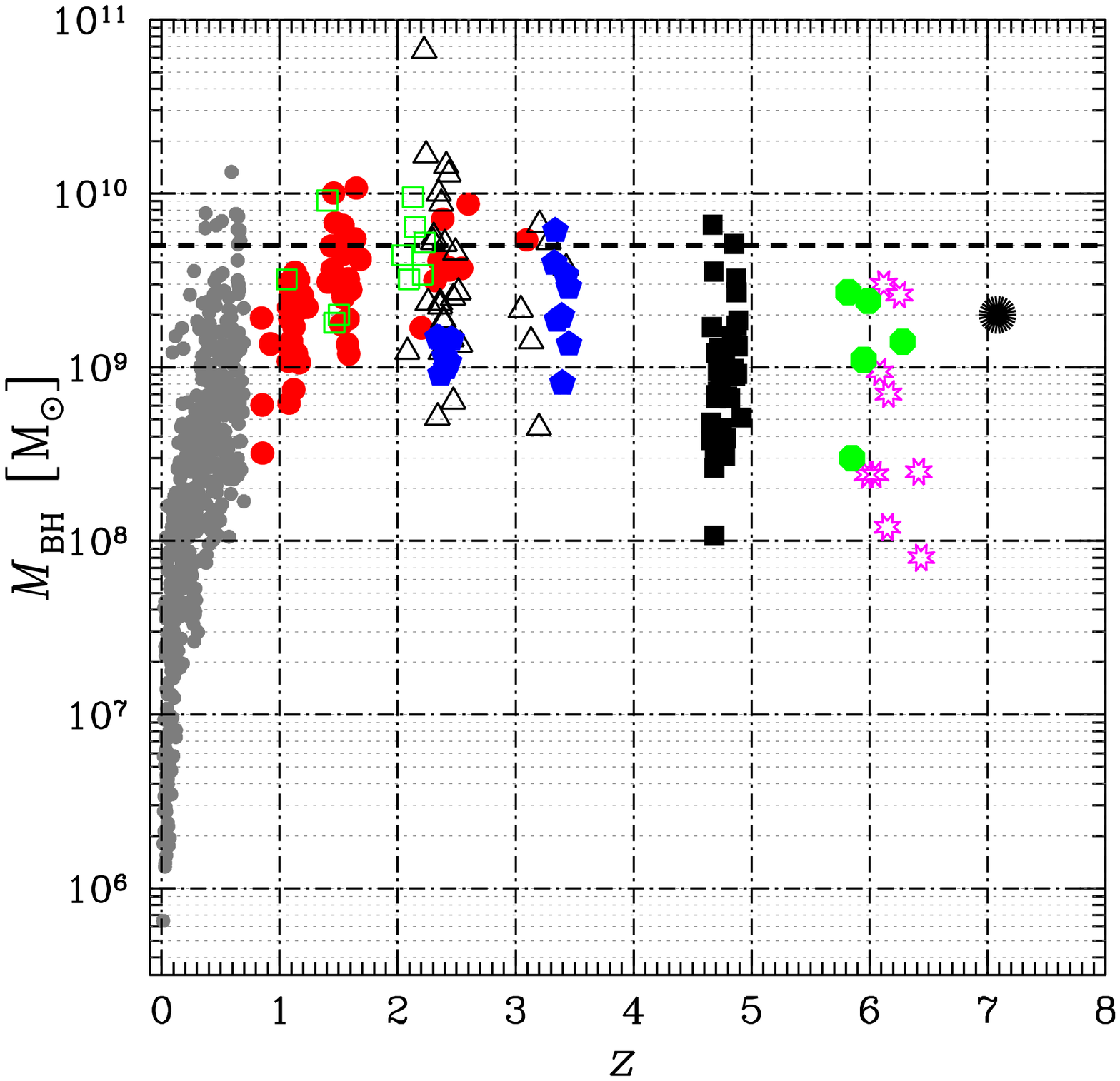}
 \caption[]{\mbh\ versus $z$\ for a low-$z$ sample \citep[grey dots][]{zamfiretal10}, and several imtermediate to high $z$ samples. Red circles: \citet{marzianietal09}; open squares: \citet{dietrichetal09}; open triangles: \citet{shemmeretal04}; filled pentagons: \citet{netzeretal07}; filled squares: \citet{trakhtenbrotetal11}; open starred octagons: \citet{willottetal10}; filled octagons: \citet{kurketal07}; large spot at $z \approx$ 7: the high-$z$ quasar whose discovery was announced in late June 2011 \citep{mortlocketal11}. The dashed line marks \mbh\  = 5$ \cdot 10^{9}$ \msol. }\label{fig:massz} 
 \end{figure}

\section{Toward higher and higher redshifts: the evidence for  a turnover} 

FWHM H$\beta$ measures can provide good \mbh\  estimates out to $z \approx$ 3.8. 8m-class telescopes yield high s/n measures for the high luminosity tail of the quasar optical luminosity function. Sampling over a wider part of the optical luminosity function (down to Seyfert 1 luminosities ($m_\mathrm{V}\approx$-23) becomes possible with the next generation of large telescopes. BH mass estimates beyond $z \approx$ 3.8 or simply larger samples beyond $z \approx 0.7$\ requires use of \hb\ surrogate lines. The two most used surrogates are \mgii\  and \civ. \civ\ cannot be trusted but \mgii\ may be able to serve as a surrogate virial estimator for the highest redshift quasars currently known. Spectra can be obtained for $z \approx 6$ quasars in the K band \citep{kurketal07}.

The best \mbh\ estimates out to $z \approx$ 4 show no evidence of a turnover which would reflect the epoch when the largest black holes were still growing. Instead we see constant \mbh\ upper limit near $\log$ \mbh\ $\approx9.7$  if we trust in part (which we do not) the measurements based on \civ\ \citep{shenetal08,labitaetal09}.  There may be a change at higher redshift \citep{trakhtenbrotetal11} if  we consider only \mbh\ values obtained from \hb\ and \mgii\ measures. Fig. \ref{fig:massz}  combines a low- $z$ sample \citep{zamfiretal10} with samples using \hb\ observed with IR spectrometers (to $z \approx 3.5$) and \mgii\ using optical and IR measures. We see a possible turnover in estimated \mbh\ at the highest redshifts although some care is needed in interpreting these results. Most sources in the range $1 \simlt z \simlt 2.5$\ were selected from the brightest  quasars in the  Hamburg ESO survey and are the most luminous quasars known in that redshift range. Observations at very high redshift refer to much fainter quasars. We must restrict our attention to the high-end of the mass distribution when we evaluate the significance of the turnover. Counting sources with masses in the ranges $9.25 \le \log$ \mbh $\le$ 9.75 and $8.75 \le \log$ \mbh $\le$ 9.25, we find that the ratio of the numbers of less-massive to more-massive sources  at redshift $\simgt$ 4 is  lower than for the samples at $1 \simlt z \simlt 3.8$. A simple application of Poisson statistics to these ratios confirms a real trend.  Given the different sample selection criteria at different redshifts we believe that more data are needed before the turnover can be regarded as established.  

As a final consideration we note that the computed \mbh\ may not be critical for concordance cosmology, since black holes can grow to the observed masses in a duty cycle that is significantly shorter than the age of the Universe at $z\approx$ 6 according to \citet{trakhtenbrotetal11}.

% Along with the low-$z$\ sample, the HE objects show the neat trend in the minimum detectable mass expected for flux-limited samples (if no super-Eddington radiators are possible). Together with the more limited numbers, they create the appearance of a shallow decrease that should we taken with some care. 

\section{Conclusion} 

\mbh\ computation techniques for large samples of quasars are rough and the lack of accuracy in \mbh\ estimates is serious. There are several areas that could lead to significant improvement:
\begin{itemize}
\item  a significant reduction in scatter could be achieved by more careful selection of virial broadening estimators (best are H$\beta$ and MgII); 
\item a second factor is related to knowledge of the BLR structure that is still hotly debated \citep{gaskell09b}. There is evidence that Pop. A and B sources show different BLR structure and kinematics. Significantly different $f$\ values are likely associated with the two populations;
\item photoionization methods should be favored over methods based on the \rb-$L$ correlation.  
\end{itemize}
Considering the large scatter introduced by uncertainties in the factors entering the virial relation it is still not surprising that \mbh\ estimates and those derived by randomly reassigning the quasar broad-line widths to different objects show such similarities in the  \mbh\  vs. $z$  plane \citep{croom11}. However this provocative result may not stand for long.
 
\bigskip 
 {The work was presented as an invited talk at special workshop "Spectral lines
and super-massive black holes" held on June 10, 2011 as a part of activity in
the frame of COST action 0905 "Black holes in an violent universe" and as a
part of the 8$^\mathrm{th}$ Serbian Conference on Spectral Line Shapes in Astrophysics. We are indebted to Martin Gaskell for discussions and many insightful suggestions.
 We also acknowledge with gratitude the hospitality and good organization of the  Conference in Div\v{c}ibare (Luka, Dragana, Darko and all the others of the organizing committee). 
%  he made during the   meeting in . 
%and to an anonymous referee who pointed out many valuable papers of Martin Gaskell we were not familiar with. 
}

 \vfill

 \bibliographystyle{model2-names}
\bibliography{biblio.bib}

\end{document}